\documentclass{article}
\pdfoutput=1 

\usepackage{mathtools, indentfirst}
\usepackage{booktabs}
\usepackage[english]{babel}
\usepackage{amsmath,amssymb,amsbsy,amstext, amsthm, simplewick,
  amsfonts}
\usepackage{graphicx}
\usepackage{upgreek}
\usepackage{framed}
\usepackage{wrapfig}
\usepackage{multirow}
\usepackage{bbm}
\usepackage[numbers,sort&compress]{natbib}
\usepackage[svgnames,dvipsnames,x11names]{xcolor}
\usepackage[utf8x]{inputenc}
\usepackage{bm}
\usepackage{float}
\usepackage{geometry}
\usepackage{yfonts}
\usepackage{caption}
\usepackage{subcaption}
\usepackage{sidecap}
\usepackage{longtable}
\usepackage{anyfontsize}

\usepackage{epstopdf}
\usepackage{cancel}
\usepackage{tcolorbox}

\def\xyma{\xymatrix@M.7em}
\def\xymas{\xymatrix@M.1em}

\newcommand{\Comment}[1]{{}}
\definecolor{darkblue}{rgb}{0.15,0.35,0.55}
\definecolor{reddish}{rgb}{0.65, 0.2, 0.2}
\definecolor{darkgreen}{RGB}{50,150,0}
\definecolor{greyish2}{rgb}{.96,.96,.96}

\usepackage{pgfornament}

\usepackage[linktocpage=true]{hyperref}

\usepackage{orcidlink}

\hypersetup{
colorlinks=true,
citecolor=darkblue,
linkcolor=reddish,
urlcolor=darkblue,
pdfauthor={},
pdftitle={},
pdfsubject={}
}

\usepackage[nameinlink]{cleveref}

\DeclareFontFamily{OT1}{rsfs10}{}
\DeclareFontShape{OT1}{rsfs10}{m}{n}{ <-> rsfs10 }{}
\DeclareMathAlphabet{\mathscript}{OT1}{rsfs10}{m}{n}

\def\gsim{ \lower .75ex \hbox{$\sim$} \llap{\raise .27ex \hbox{$>$}} }
\def\lsim{ \lower .75ex \hbox{$\sim$} \llap{\raise .27ex \hbox{$<$}} }

\usepackage{latexsym,amsmath,amssymb,epsfig}

 \usepackage{tikz}
\usetikzlibrary{decorations}
\pgfdeclaredecoration{complete sines}{initial}
{
    \state{initial}[
        width=+0pt,
        next state=upsine,
        persistent precomputation={\pgfmathsetmacro\matchinglength{
            \pgfdecoratedinputsegmentlength / int(\pgfdecoratedinputsegmentlength/\pgfdecorationsegmentlength)}
            \setlength{\pgfdecorationsegmentlength}{\matchinglength pt}
        }] {}
    \state{upsine}[width=\pgfdecorationsegmentlength,next state=downsine]{
        \pgfpathsine{\pgfpoint{0.25\pgfdecorationsegmentlength}{0.5\pgfdecorationsegmentamplitude}}
        \pgfpathcosine{\pgfpoint{0.25\pgfdecorationsegmentlength}{-0.5\pgfdecorationsegmentamplitude}}
    }
    \state{downsine}[width=\pgfdecorationsegmentlength,next state=upsine]{
        \pgfpathsine{\pgfpoint{0.25\pgfdecorationsegmentlength}{-0.5\pgfdecorationsegmentamplitude}}
        \pgfpathcosine{\pgfpoint{0.25\pgfdecorationsegmentlength}{0.5\pgfdecorationsegmentamplitude}}
}
    \state{final}{}
}

\definecolor{greyish}{rgb}{.90,.90,.90}
\definecolor{greyish2}{rgb}{.96,.96,.96}
\usepackage{xcolor,colortbl}
\usepackage{tcolorbox}

\usepackage[all]{xy}

\usepackage{ytableau}

\numberwithin{equation}{section}

\definecolor{turquoise}{RGB}{0,206,209}

\usepackage{physics,enumitem}
\usepackage[mathscr]{eucal}
\usepackage[normalem]{ulem}

\begin{document}
\renewcommand{\thefootnote}{\fnsymbol{footnote}}
\vspace{0truecm}
\thispagestyle{empty}

\vspace*{-0.3cm}

\begin{center}
{\fontsize{21}{18} \bf Geometric Symmetries for the Vanishing of}\\[14pt]
{\fontsize{21}{18} \bf the Black Hole Tidal Love Numbers}
\end{center}

\vspace{.15truecm}

\begin{center}
{\fontsize{13}{18}\selectfont
Roman Berens \orcidlink{0000-0003-1509-5463},${}^{\rm a}$\footnote{\texttt{roman.berens@vanderbilt.edu}} 
Lam Hui \orcidlink{0000-0001-7003-4132},${}^{\rm b}$\footnote{\texttt{lh399@columbia.edu}} 
Daniel McLoughlin \orcidlink{0009-0005-3535-2334},${}^{\rm b}$\footnote{\texttt{dcm2183@columbia.edu}} 
\\
  \vspace{.3cm}
Riccardo Penco \orcidlink{0000-0002-6735-4319},${}^{\rm c}$\footnote{\texttt{rpenco@andrew.cmu.edu}}
and 
John Staunton \orcidlink{0009-0004-1661-9577}${}^{\rm b}$\footnote{\texttt{j.staunton@columbia.edu}}
}
\end{center}
\vspace{.4truecm}

\centerline{{\it ${}^{\rm a}$Department of Physics \& Astronomy,
    Vanderbilt University, Nashville, TN 37212, U.S.A.}}
  
\vspace{.3cm}

\centerline{{\it ${}^{\rm b}$Center for Theoretical Physics, Department of Physics,}}
\centerline{{\it Columbia University, New York, NY 10027, U.S.A.}}

\vspace{.3cm}

\centerline{{\it ${}^{\rm c}$Department of Physics, Carnegie Mellon
    University, Pittsburgh, PA 15213, U.S.A.}} 
 
\vspace{.25cm}

\vspace{.3cm}

\begin{abstract}

\noindent We present a unified geometric perspective on the symmetries underlying the spin 0, 1 and 2 static perturbations around a Schwarzschild black hole. In all cases, the symmetries are exact, each forming an SO(3,1) group. They can be formulated at the level of the action, provided the appropriate field variables are chosen. For spin 1 and 2, the convenient variables are certain combinations of the gauge fields for even perturbations, and dual scalars for odd perturbations. The even and odd sectors each have its own SO(3,1) symmetry. In addition, there is an SO(2) symmetry connecting them, furnishing an economical description of Chandrasekhar's duality. When decomposed into spherical harmonics, the perturbations form a non-trivial representation of SO(3,1), giving rise to ladder symmetries which explain the vanishing of the tidal Love numbers. Our work builds on earlier discussions of ladder symmetries, which were formulated in terms of the Newman-Penrose scalar at the level of the Teukolsky equation. Our formulation makes it possible to state the symmetries responsible for the vanishing of the Wilson coefficients characterizing the spin 0, 1 and 2 static tidal response in the effective point particle description of a black hole.

\end{abstract}

\newpage

\setcounter{tocdepth}{2}
\tableofcontents
\newpage
\renewcommand*{\thefootnote}{\arabic{footnote}}
\setcounter{footnote}{0}

\section{Introduction}
Black hole perturbation theory is an old subject \cite{Regge:1957,Zerilli:1970se,Teukolsky:1973ha,Chandrasekhar:1985}, yet it has surprising riches, some of which were not appreciated until recently. One example has to do with the black hole Love numbers, which characterize a black hole's static tidal response. It has been well established for some time that they vanish for linear black hole perturbations in four dimensions \cite{Fang:2005qq,Damour:2009vw,Binnington:2009bb,Kol:2011vg,LeTiec:2020bos,LeTiec:2020spy,Chia:2020yla,Hui2021,Charalambous:2021mea}. Recently, progress was made in understanding the symmetries behind this phenomenon. This includes symmetries in the near zone approximation(s) \cite{Charalambous:2021kcz,Charalambous:2022,Hui2022a}, as well as exact symmetries in the static regime, known as ladder symmetries \cite{Hui2022,Berens2023,BenAchour:2022uqo,Rai:2024lho}. Other discussions on symmetries are present in \cite{Katagiri:2022, Chakraborty:2025, Bhattacharyya:2025, Lupsasca2025, Parra-Martinez:2025}. Understanding the symmetries is not only aesthetically appealing; it also has practical benefits. For instance, it appears the symmetries uncovered in linear theory extend in some form into the nonlinear regime, enabling the deduction of vanishing nonlinear black hole Love numbers \cite{Kehagias:2024rtz,Combaluzier-Szteinsznaider:2024sgb,Gounis:2024hcm}.

In the work of \cite{Hui2022,Berens2023}, it was pointed out that the ladder symmetries have a geometric origin in the case of spin 0 perturbations. The key observation is that the action for a generic scalar in a curved background is not only invariant under symmetries generated by Killing vectors (KVs), but also by a subset of additional conformal Killing vectors (CKVs) referred to as \textit{melodic} CKVs. Our main task in this paper is to show how the same story extends to spin 1 and 2 perturbations around Schwarzschild black holes. As we will see, a rather pleasing picture emerges: the SO(3,1) group generated by these melodic CKVs figures in all static linear perturbations around a Schwarzschild black hole, whether they be spin 0, 1 or 2. Moreover, in the case of spin 1 and 2, the respective even and odd sectors each have its own SO(3,1) symmetry, with an SO(2) symmetry connecting the two sectors. The ladder symmetries of \cite{Hui2022,Berens2023} follow from the representation theory of SO(3,1), and are responsible for the vanishing of the Love numbers. An additional bonus of our investigation is that the symmetries can be stated at the level of the gauge fields (or certain dual scalars), and therefore the action. This is in contrast with the earlier work of \cite{Hui2022} where the symmetries (for spin 1 and 2 perturbations) were phrased at the level of the Newman-Penrose scalar satisfying the Teukolsky equation of motion.

These symmetries are also relevant for the worldline effective field theory (EFT) of a point object \cite{Goldberger:2004jt}. Recall that the idea of a worldline EFT is to zoom out, so that an object like a star or a black hole is treated as a point particle of some mass, and finite size effects such as tidal deformability are encoded in operators localized on the particle worldline. The Wilson coefficients of such operators are the Love numbers. In \cite{Hui2022}, it was pointed out that the large $r$ limit of the (``boost" part of) SO(3,1) prohibits the appearance of Love number worldline operators for spin 0 perturbations. In this work, we will demonstrate the same applies for spin 1 and 2 perturbations.

Our paper is organized as follows. \Cref{s:StaticScalar} serves as a review of how the spin 0 case works, including a discussion of the geometric and ladder symmetries, and how they relate to the vanishing of the Love numbers. In \Cref{s:SchwarzschildVector} and \Cref{s:Spin2}, we show how the same logic applies to the spin 1 and 2 perturbations, once the appropriate field variables are chosen. We conclude with a discussion in \Cref{discuss}. Technical details are relegated to several Appendices. In particular, in \autoref{MCKV}, we derive the condition under which two conformally related metrics share the same melodic CKVs. We also work out the Noether current corresponding to the SO(3,1) symmetries.

\textit{Notation and Convention:} We use the mostly plus signature and natural units where $c = \hbar = G = 1$. Lorentz indices are labeled with Greek indices $\mu, \nu, \rho$, etc., while spatial indices are labeled with Latin indices $i, j, k$, etc. In \Cref{s:Spin2}, we will perform a decomposition into radial and angular pieces and label angular indices with $a$, $b$, $c$, etc. These angular indices are raised and lowered with the metric on the 2-sphere.

\section{Review: Static Spin 0 Perturbations}
\label{s:StaticScalar}
In this section we give a brief review of the geometric symmetries, forming an SO(3,1) group, enjoyed by static (massless) scalar perturbations around a Schwarzschild black hole. We describe how the representation theory of SO(3,1) gives rise to ladder symmetries, from which the vanishing of Love numbers follows. We also review how the large distance limit of the geometric symmetries can be applied to the worldline effective field theory of a black hole, and how it explains the vanishing of the Wilson coefficients corresponding to Love numbers. The discussion in this section closely follows that of \cite{Hui2022} and \cite{Berens2023}. 

To that end, consider the action of a free minimally coupled, massless scalar field in a Schwarzschild background
\begin{equation}
S = -\frac{1}{2}\int \dd[4]{x} \sqrt{-g} g^{\mu\nu}\partial_\mu \phi \partial_\nu \phi,
\end{equation}
where, in Schwarzschild coordinates, the metric is described by the line element
\begin{equation}
\label{eq:SchwarzschildLine}
\dd{s}^2 = -\frac{\Delta}{r^2}\dd{t}^2 + \frac{r^2}{\Delta}\dd{r}^2 + r^2\left(\dd{\theta}^2 + \sin^2\theta\dd{\varphi}^2\right).
\end{equation}
Here, $\Delta \equiv r\left(r - r_s\right)$, where $r_s$ is the Schwarzschild radius.

\subsection{Symmetries of the Effective Metric}
\label{symmetriesSpin0}

We are interested in static perturbations. With time derivatives set to zero, the scalar $\phi$ can be thought of as effectively living in a three-dimensional (3D) space with the metric $\hat g_{ij}$, satisfying
\begin{equation}
\sqrt{\hat{g}} \hat{g}^{ij} = \sqrt{-g} g^{ij}.
\end{equation}
The action thus takes the form:
\begin{equation}
\label{eq:MinimallyCoupledSpin0}
S = -\frac{1}{2}\int \dd{t} \dd[3]{x} \sqrt{\hat{g}} \hat{g}^{ij} \partial_i \phi \partial_j \phi.
\end{equation}
Note that, due to the presence of the determinant, the 3D metric $\hat{g}_{ij}$ is not equal to the spatial components of the full four-dimensional metric $g_{\mu\nu}$. This is an algebraic tensor relationship which can be solved exactly, yielding
\begin{equation}
\label{eq:GeneralEffMetric}
\hat{g}_{ij} = -g_{tt}\left(g_{ij} - \frac{g_{ti}g_{tj}}{g_{tt}}\right).
\end{equation}
This is the effective metric for a static scalar in a general spacetime. In the case of the Schwarzschild background, the associated effective 3D metric is given by:
\begin{equation}
\label{eq:3DSchwarzschildScalarEffectiveMetric}
\dd{\hat{s}}^2 = \dd{r}^2 + \Delta\left(\dd{\theta}^2 + \sin^2\theta \dd{\varphi}^2\right).
\end{equation}
As far as isometries go, this metric has the expected 3 rotational Killing vectors (KVs) as well as 7 conformal Killing vectors (CKVs). Our scalar $\phi$ is {\it not} conformally coupled (and the Ricci scalar corresponding to $\hat g_{ij}$ does not vanish), which normally means the 7 CKVs do not generate symmetries. However, it turns out 3 of them obey a special condition which guarantees that they do. We appeal to the following theorem, which is proven in \cite{Berens2023} and elaborated on in \autoref{MCKV}.\footnote{The integration over $t$ in \autoref{eq:MinimallyCoupledSpin0} is trivial in our application, and is neglected in the statement of our theorem.}

\begin{framed}
\noindent {\it Theorem.} For a scalar $\phi$ in $d$ dimensions with the following action (here we generalize the discussion to consider the metric $\hat g_{ij}$ in $d$ dimensions):
\begin{equation}
S = -\frac{1}{2} \int \dd[d]{x} \sqrt{\abs{\hat{g}}} \left(\hat{g}^{ij} \partial_i \phi \partial_j \phi + \alpha \hat{R} \phi^2\right), 
\end{equation}
where $\hat R$ is the Ricci scalar corresponding to the metric $\hat{g}_{ij}$ and $\alpha$ is a constant, a CKV $\xi$ generates a symmetry provided it satisfies the {\it melodic condition}:
\begin{equation}
\Box \nabla_i \xi^i = 0 \, ,
\end{equation}
where the covariant derivatives are defined with respect to the metric $\hat g_{ij}$. The corresponding symmetry is
\begin{equation}
\label{deltaxiphi}
\delta_\xi \phi = \xi^i \nabla_i \phi + \frac{d - 2}{2d} \left(\nabla_i \xi^i\right) \phi\, .
\end{equation}
Note that this holds for arbitrary values of $\alpha$, including zero. In other words, it need not take the conformal value $(d - 2) / [4(d - 1)]$. Additionally, note that a KV automatically satisfies the melodic condition
by virtue of $\nabla_i \xi^i = 0$.

\noindent {\it Corollary.} It can be shown that $\xi^i$ is a melodic CKV for the metric $\hat{g}_{ij}$ if and only if it is a KV for the following conformally related metric $\tilde{g}_{ij}$:
\begin{equation}
\label{eq:rescaledmetric}
\tilde{g}_{ij} = \hat{R} L_0^2 \hat g_{ij}, 
\end{equation}
where $\hat{R}$ is the Ricci scalar of the metric $\hat{g}_{ij}$, and $L_0$ is an arbitrary length scale to fix the dimensions.
\end{framed}

That we can find all melodic CKVs from studying the KVs of a rescaled metric is particularly useful in our static three-dimensional case in which there is a known procedure for finding the exact number of Killing vectors \cite{Kruglikov2018}. In the case of the Schwarzschild effective metric in \autoref{eq:3DSchwarzschildScalarEffectiveMetric}, the Ricci scalar is
\begin{equation}
\label{eq:StaticScalarRicci}
\hat{R}_{\text{scalar}} = \frac{r_s^2}{2\Delta^2}.
\end{equation}
Observe that the conformally rescaled metric,
\begin{equation}
\label{eq:EAdS}
\dd{\tilde{{s}}}^{\:2} = \frac{r_s^2 L_0^2}{2\Delta^2}\left(\dd{r}^2 +
  \Delta \left(\dd{\theta}^2 + \sin^2\theta
    \dd{\varphi}^2\right)\right) ,
\end{equation}
describes a Euclidean anti-de Sitter space (EAdS), as demonstrated by the fact that the Ricci tensor for the metric in \autoref{eq:EAdS} is $\tilde{{R}}_{ij} = \tilde{{R}} \tilde{{g}}_{ij} /3$. This space, also known as hyperbolic space, has six KVs. Three of them are the familiar rotational KVs. The other three can be thought of as translations in the hyperbolic space, or boosts in the embedding space.\footnote{A discussion on how to find the Killing vectors is presented in \cite[Appendix A]{Hui2021}. To summarize, EAdS can be embedded in flat space in one higher dimension: $\dd{s}^2 = -\dd{x}_0^2 + \dd{x}_1^2 + \dd{x}_2^2 + \dd{x}_3^2$, subject to the constraint $-x_0^2 + x_1^2 + x_2^2 + x_3^2 = 2L_0^2$. The metric \autoref{eq:EAdS} can be obtained by the mapping: 
\begin{equation}
x_1 = \frac{r_s L_0}{\sqrt{2\Delta}}\sin\theta\cos\varphi, \qquad x_2
= \frac{r_s L_0}{\sqrt{2\Delta}}\sin\theta \sin\varphi, \qquad
\text{and} \qquad x_3 = \frac{r_s L_0}{\sqrt{2\Delta}}\cos\theta \, .
\end{equation}
The three rotations and three boosts in the flat 4D embedding space then map to the six KVs of the EAdS space.} By the Corollary above, they are thus melodic CKVs of the pre-rescaled metric, i.e. the effective 3D metric \autoref{eq:3DSchwarzschildScalarEffectiveMetric}, and by the Theorem they generate symmetries for the scalar $\phi$. The form of the melodic CKVs, including the three rotations and three ``boosts'', are:
\begin{subequations}
    \begin{align}
    \label{eq:AllMCKVs1}
        iK_1 &= -\frac{2\Delta}{r_s}\sin\theta\cos\varphi\,\partial_r + \frac{\Delta'}{r_s}\left(\cos\theta\cos\varphi\,\partial_\theta - \csc\theta\sin\varphi \,\partial_\varphi\right),\\
        iK_2 &= -\frac{2\Delta}{r_s}\sin\theta\sin\varphi \,\partial_r + \frac{\Delta'}{r_s}\left(\cos\theta \sin\varphi \,\partial_\theta + \csc\theta \cos\varphi\,\partial_\varphi\right),\\
        \label{eq:AllMCKVs3}
        iK_3 &= -\frac{2\Delta}{r_s}\cos\theta \,\partial_r - \frac{\Delta'}{r_s}\sin\theta\,\partial_\theta,\\
        \label{eq:AllMCKVs4}
        iJ_1 &= -\sin\varphi \,\partial_\theta - \cot\theta \cos\varphi \,\partial_\varphi,\\
        iJ_2 &= \cos\varphi \,\partial_\theta - \cot\theta \sin\varphi \,\partial_\varphi,\\
        iJ_3 &= \partial_\varphi.
        \label{eq:AllMCKVs6}
    \end{align}
\end{subequations}
The three rotation generators exactly agree with
\cite[eq. (2.3)]{Berens2023}, while the boost generators agree up to an irrelevant normalization, $-2/r_s$. This choice of normalization reproduces the standard SO(3,1) Lie algebra
\begin{equation}
\label{eq:MCKVsLA}
\comm{K_i}{K_j} = -i \epsilon_{ij}^{\hphantom{ij}k} J_k, \qquad \comm{K_i}{J_j} = i\epsilon_{ij}^{\hphantom{ij}k}  K_k \qquad \text{and} \qquad \comm{J_i}{J_j} = i\epsilon_{ij}^{\hphantom{ij}k} J_k.
\end{equation}
Note that, in the above, $\epsilon_{ijk}$ are Levi-Civita symbols, not tensors.\footnote{\label{footnote3} If one is interested in the $r_s \rightarrow 0$ flat space limit, it is more convenient to consider $K_i^{\rm flat} \equiv r_s K_i$ instead of $K_i$ as the symmetry generator. In that case, the algebra becomes 
\begin{equation}
\comm{K_i^{\rm flat}}{K_j^{\rm flat}} = 0, \qquad \comm{K_i^{\rm flat} }{J_j} = i\epsilon_{ij}^{\hphantom{ij}k} K_k^{\rm flat}, \qquad \text{and} \qquad \comm{J_i}{J_j} = i\epsilon_{ij}^{\hphantom{ij}k} J_k.
\end{equation}
Not surprisingly, $K_i^{\rm flat}$ is the generator for conformal transformation. Thus we sometimes refer to $K_i$ as the generalized conformal transformation.}

\subsection{Implications in Harmonic Space---Vanishing of the Love Numbers}
\label{spin0harmonic}

To spell out the physical effects of these symmetries, it is helpful to decompose the scalar field $\phi$ in the basis of spherical harmonics, 
\begin{equation}
\phi\left(r, \theta, \varphi\right) = \sum_{\ell = 0}^{\infty} \sum_{m
  = -\ell}^{\ell} \phi_{\ell m}\left(r\right) Y_{\ell}^m\left(\theta,
  \varphi\right) \, .
\end{equation}
Besides leaving the action invariant up to a total derivative, A symmetry maps a solution of the equations of motion to a solution, i.e. if $\phi$ is a solution, so is $\delta_\xi\phi$, where $\xi$ is a KV or melodic CKV (\autoref{deltaxiphi}). If $\xi$ is a rotation KV, we are familiar with its effect at the level of the spherical harmonics: it mixes $\phi_{\ell m}$'s at different $m$'s (unless the rotation is around the $z$-axis, in which case no raising or lowering of $m$ occurs). As we will see below, if $\xi$ is the melodic CKV $K_3$, it mixes $\phi_{\ell m}$'s at different $\ell$'s. In other words, think of the $\phi_{\ell m}$'s lined up as a giant column vector, furnishing a representation of SO(3,1). As argued in \cite{Berens2023}, it turns out to be a principal series representation. The raising and lowering of $\ell$ gives rise to a ladder structure, as we will show below.

The equation of motion in harmonic space reads:
\begin{equation}
\label{phieom}
\left[ \Delta \partial_r (\Delta \partial_r) - \ell(\ell+1) \right] \phi_{\ell} (r) = 0 \, .
\end{equation}
The spherical symmetry of the Schwarzschild metric guarantees the equation is independent of $m$. This is why, from now on, we drop the $m$ index of $\phi_{\ell m} (r)$, i.e. $\phi_{\ell m} (r)$ can still depend on $m$ if one wishes, but its $r$-dependence is independent of $m$. 

Let us focus on the transformation effected by $K_3$ (the effects of $K_1$ and $K_2$ follow from that of $K_3$, commuted with a rotation, i.e. $K_1$ and $K_2$ raise and lower $m$ in addition to $\ell$). Consider the transformation ${\xi}^i \partial_i$, which is a rescaled version of $K_3$ (chosen for later convenience):
\begin{subequations}
\begin{align}
\label{eq:MainMCKV}
\xi^i \partial_i &\equiv -\frac{ir_s}{2} K_3,\\
&= \Delta \cos\theta \,\partial_r + \frac{\Delta'}{2} \sin\theta\,\partial_\theta.
\end{align}
\end{subequations}
The variation of a scalar field with respect to this CKV, \autoref{deltaxiphi}, is given by
\begin{equation}
\label{eq:Scalarvarphi}
    \delta_\xi \phi = \Delta \cos\theta \,\partial_r\phi +
    \frac{\Delta'}{2}\partial_\theta\left(\sin\theta\,\partial_\theta\phi\right)
    \, .
\end{equation}
If we plug in the expansion into spherical harmonics, without assuming $\phi$ is a solution to the equations of motion, we find\footnote{As pointed out in \cite{Berens2023}, this is most easily derived using
\begin{subequations}
\begin{align}
\sin\theta\,\partial_\theta Y_{\ell}^m &= \ell f\left(\ell\right) Y_{\ell + 1}^m - \left(\ell + 1\right) f\left(\ell - 1\right) Y_{\ell - 1}^m,\\
\cos\theta Y_{\ell}^m &= f\left(\ell\right) Y_{\ell + 1}^m + f\left(\ell - 1\right) Y_{\ell - 1}^m.
\end{align}
\end{subequations}}
\begin{equation}
\label{eq:LadderfromK3}
\delta_{{\xi}}\left(\phi_{\ell} Y_{\ell}^m\right) = -f\left(\ell\right) D_{\ell}^+\phi_{\ell} Y_{\ell + 1}^m + f\left(\ell - 1\right) D_{\ell}^- \phi_{\ell} Y_{\ell - 1}^m,
\end{equation}
where 
\begin{equation}
f\left(\ell\right) = \sqrt{\frac{\left(\ell - m + 1\right)\left(\ell + m + 1\right)}{\left(2\ell + 1\right)\left(2\ell + 3\right)}},
\end{equation}
and the operators $D_{\ell}^{\pm}$ are defined by 
\begin{equation}
\label{eq:Spin0Ladders}
D_{\ell}^+ \equiv -\Delta \partial_r - \frac{\ell + 1}{2} \Delta' \qquad \text{and} \qquad D_{\ell}^- \equiv \Delta \partial_r - \frac{\ell}{2}\Delta'.
\end{equation}
The operators $D_{\ell}^{\pm}$ have the interpretation of being ladder operators. Since the first term in $\delta_{\xi}\left(\phi_{\ell} Y_{\ell}^m\right)$ multiplies $Y_{\ell + 1}^m$, $D_{\ell}^+\phi_{\ell}$ is a solution to the radial equation of motion at level $\ell + 1$ if $\phi_\ell$ is a solution at level $\ell$. Similarly, $D_{\ell}^- \phi_{\ell}$ is a solution to the radial equation of motion at level $\ell - 1$.

To better understand the implications of $D_{\ell}^{\pm}$ as ladder operators, it is helpful to review how they appear at the level of the radial equation of motion. Rewriting \autoref{phieom} as
\begin{equation}
\label{phieom2}
H_{\ell} \phi_{\ell} = 0 \qquad \text{where} \qquad H_{\ell} \equiv
-\Delta\left(\partial_r\left(\Delta \partial_r\right) - \ell\left(\ell
    + 1\right)\right) \, ,
\end{equation}
where $H_\ell$ can be thought of as a Hamiltonian operator,
and utilizing the intertwining relation
\begin{equation}
\label{eq:MainIntertwiningRelation}
D_{\ell + 1}^- D_{\ell}^+ - D_{\ell - 1}^+ D_{\ell}^- = \frac{2\ell + 1}{4}r_s^2,
\end{equation}
one can show that the Hamiltonian is expressible in terms of the ladder operators
\begin{equation}
\label{HDD}
H_{\ell} = D_{\ell + 1}^- D_{\ell}^+ - \frac{\left(\ell + 1\right)^2}{4} r_s^2 = D_{\ell - 1}^+ D_{\ell}^- - \frac{\ell^2}{4}r_s^2.
\end{equation}
The relationships between the Hamiltonian and the ladder operators that justify the interpretation of $D_{\ell}^{\pm}$ as ladder operators are
\begin{equation}
\label{eq:LadderAlgebra}
H_{\ell + 1} D_{\ell}^+ = D_{\ell}^+ H_{\ell} \quad , \quad  H_{\ell - 1} D_{\ell}^- = D_{\ell}^- H_{\ell}.
\end{equation}
These show that $D_{\ell}^+ \phi_{\ell}$ is a solution to the equation of motion at level $\ell + 1$ so long as $\phi_{\ell}$ is a solution at level $\ell$. Similarly, $D_{\ell}^- \phi_{\ell}$ is a solution at level $\ell - 1$ so long as $\phi_{\ell}$ is a solution at level $\ell$. 

How is the ladder structure helpful in understanding the vanishing of the Love numbers? Let us first recall what it is about the vanishing of the Love numbers that seems to demand an explanation. Looking at the equation of motion \autoref{phieom2}, which is second order, it is clear at large $r$, the most general solution should be a superposition of two independent large $r$ asymptotics: $r^\ell$ and $1/r^{\ell+1}$. On the other hand, as $r$ approaches the horizon $r_s$, the most general solution should be a superposition of the two independent asymptotics: a constant and ${\rm log\,}(1 - r_s/r)$. The physical solution should be one that's regular at the horizon, and therefore approaches a constant there. Generically, one might expect a solution that approaches a constant at the horizon to be a superposition of $r^\ell$ and $1/r^{\ell+1}$ at large $r$. The vanishing of the Love numbers violates this expectation: it turns out the solution that is regular at the horizon has $r^\ell$ behavior, but no $1/r^{\ell+1}$ tidal tail at all, at large $r$. Interpreting $r^\ell$ as an external tidal field, and $1/r^{\ell+1}$ as the tidal response, leads to the conclusion that the tidal deformability is zero. The question of why the Love numbers vanish is equivalent to asking why the generic expectation fails.

We address this question in two steps. First, let's study the bottom rung of the ladder: $\ell = 0$. The equation of motion takes the form $D^+_{-1} D^-_0 \phi_0 = 0$ (\autoref{HDD}). Taking a cue from the simple harmonic oscillator, let us postulate that the desired (regular) solution satisfies:
\begin{equation}
D_0^- \phi_0 = 0.
\end{equation}
This is analogous to saying that the $\ell = 0$ {\it ground state} should be annihilated by the lowering operator. This turns what was originally a second order equation into a first order one, for which at $r \rightarrow r_s$ or $r \rightarrow \infty$, there can only be a single asymptotic behavior respectively. In fact $D_0^- = \Delta \partial_r$ is sufficiently simple that we can see the solution is $\phi_0=$ constant, i.e. it has the correct regular behavior at the horizon, {\it and} it has no $1/r$ tidal tail at large $r$ (i.e. it can't possibly be a superposition of $r^0$ and $1/r$ anyway, because $D_0^- \phi_0 = 0$ is a first order equation).

The second step in understanding the vanishing of the Love numbers is to appeal to the ladder structure. Starting from the ground state $\phi_0 =$ constant solution, the solution at any level $\ell$ can be built by successively climbing the ladder:
\begin{equation}
\phi_{\ell} \propto D_{\ell - 1}^+ \dots D_1^+ D_0^+ \phi_0.
\end{equation}
The form of the raising operators (\autoref{eq:Spin0Ladders}) makes it clear that $\phi_{\ell}$ is regular at the horizon, and grows as $r^\ell$, without the $1/r^{\ell+1}$ tail, at large $r$. To summarize, the vanishing of the Love numbers requires both a ladder structure and the existence of a ground state. The ground state condition reduces a second-order problem to a first-order one, and the ladder ensures that the solutions at level $\ell \geq 0$ are both regular at the horizon and purely growing at infinity, meaning the Love numbers vanish for all integer $\ell \geq 0$.

Let us close this section by making connection with the ladder symmetries pointed out by \cite{Hui2022}. From \autoref{eq:LadderfromK3}, we see that one can obtain a level $\ell-1$ solution from a level $\ell$ solution by $\delta \phi_{\ell-1} = f(\ell-1) D^-_\ell \phi_\ell$. Employing the same equation with $\ell \rightarrow \ell-1$, we see that one can also obtain a level $\ell$ solution from a level $\ell-1$ solution by $\delta \phi_\ell = - f(\ell-1) D^+_{\ell-1} \phi_{\ell-1}$. With this understanding, if we isolate the two levels $\ell-1$ and $\ell$, it can be shown that the scalar action in harmonic space is invariant under
\begin{equation}
\delta\phi_\ell = - D^+_{\ell-1} \phi_{\ell-1} \quad , \quad
\delta\phi_{\ell-1} = D^-_\ell \phi_\ell \, .
\end{equation}
This has the flavor of a supersymmetry, and can be formulated as such \cite{Hui2022}. This is called a \textit{vertical ladder symmetry}. There is a separate symmetry that also makes use of the ladder structure. Looking at the \autoref{phieom}, it's clear that for $\ell=0$, if $\phi_0$ is a solution, so is $\Delta\partial_r \phi_0$. In other words, $\delta\phi_0 = Q_0\phi_0$ is a symmetry, with $Q_0 \equiv \Delta\partial_r$. The ladder structure can then be used to compose a symmetry at level $\ell$, i.e. $\delta\phi_\ell = Q_\ell \phi_\ell$ is a symmetry, with $Q_\ell \equiv D^+_{\ell-1} D^+_{\ell-2} ... D_0^+ Q_0 D^-_1 ... D^-_{\ell-1} D^-_\ell$. This is a symmetry that takes a level $\ell$ solution and maps it to another level $\ell$ solution. This is called a \textit{horizontal ladder symmetry}. See \cite{Hui2022} for further discussions of how
the ladder symmetries are useful.\footnote{See also \cite{BenAchour:2022uqo} for a discussion of the connection of the horizontal ladder symmetry with the Wronskian.} It's worth emphasizing they originate from the representation theory of the geometric symmetries SO(3,1).

\subsection{Application to the Worldline EFT}
\label{eftspin0}

In this subsection, we examine the problem of the vanishing of spin 0 Love numbers from the perspective of the worldline EFT, which treats the black hole as a generic point object from the perspective of an observer far away \cite{Goldberger:2004jt}. This point object has a mass $m$, associated with which is the point-particle action:
\begin{equation}
\label{Spp}
S_{\text{pp}} = \frac{1}{2} \int \left(e^{-2} g_{\mu\nu} \dv{x^\mu}{\tau} \dv{x^\nu}{\tau} - m^2\right) e \dd{\tau}\, ,
\end{equation}
which is localized at the particle with proper time $\tau$ and worldline einbein $e$.  A complete discussion of the EFT would include a construction of the Schwarzschild background order by order in metric perturbation around Minkowski, by supplementing the above with the Einstein-Hilbert bulk action (see e.g. \cite{Ivanov:2022, Mougiakakos:2024}). We will skip this part of the discussion, and consider the large distance limit, in which the geometry is well approximated by Minkowski. We also work in the limit of negligible gravitational backreaction from the scalar $\phi$.

With this preamble, the scalar action is:
\begin{equation}
  S_{\rm scalar} = S_{\text{bulk}} + S_{\text{Love}} \, ,
\end{equation}
where
\begin{subequations}
\begin{align}
S_{\text{bulk}} &= -\frac{1}{2}\int \delta^{ij} \, \partial_i \phi \,
                  \partial_j \phi \, \dd[4]{x} ,\\
S_{\text{Love}} &= \sum_{\ell = 0}^{\infty}
                  \frac{\lambda_{\ell}}{2\ell!} \int \partial_{(i_1}
                  \dots \partial_{i_\ell)^T}\phi \,\,\, \partial^{(i_1}
                  \dots \partial^{i_\ell)^T}\phi \,\,\,  \delta^{(3)}\left(
                  {\vec{x}}
                  \right)  \dd[4]{x} .
\end{align}
\end{subequations}
Here, we have set all time derivatives to zero, because we are interested in static field configurations, and the object is assumed to be at rest at the origin. The symbol $\partial_{(i_1} \dots   \partial_{i_\ell)^T}$ represents a symmetric and traceless combination of derivatives. The corresponding scalar equation of motion is
\begin{equation}
\label{scalareomEFT}
\partial_i \partial^i \phi = \sum_{\ell = 0}^{\infty} \frac{\lambda_{\ell}}{\ell!} \partial_{(i_1} \dots \partial_{i_{\ell})^T} \left(\left(\partial^{(i_1} \dots \partial^{i_{\ell})^T} \phi\right) \delta^{(3)}\left(\vec{x}\right)\right)
\end{equation}
It can be seen that assuming an external tidal field of $\phi \sim r^\ell$ (which is zeroth-order in $\lambda_\ell$), when inserted into the above, produces a (first order) tidal response of $\phi \sim \lambda_\ell /r^{\ell+1}$. Thus, the Wilson coefficient $\lambda_\ell$ in the EFT can be taken to define the spin 0 Love number at multipole $\ell$. It captures a finite-size effect---the spin 0 tidal deformability of the object \cite{Goldberger:2004jt}. Determining $\lambda_\ell$ is a matter of matching this EFT prediction with an ultraviolet (UV) computation appropriate to the object in question. For a black hole, matching with the UV computation such as that in \Cref{spin0harmonic}  tells us $\lambda_\ell = 0$ for all $\ell$.

When encountering an object with vanishing Love numbers, a natural question arises: what symmetry forbids the existence of Love number operators in the EFT? Alternatively: if $\lambda_\ell \ne 0$, what symmetry is broken? Notice this question is phrased at the level
of the EFT for a generic point object, which might or might not be a black hole. Nonetheless, we can take a cue from the UV computation for the black hole. Consider the large $r$ limit of \autoref{eq:Scalarvarphi}, i.e.
\begin{equation}
\label{eq:scalarspecialconformal0}
\delta_\xi \phi = r^2\cos\theta \, \partial_r\phi + r\partial_\theta\left(\sin\theta \phi\right).
\end{equation}
This symmetry transformation takes a more familiar form if written in Cartesian coordinates:
\begin{equation}
\label{eq:scalarspecialconformal2}
\delta_\xi \phi = c^i (x_i - \vec x^2 \partial_i + 2 x_i \vec x \cdot \vec
\partial) \phi \, ,
\end{equation}
where the indices are raised and lowered with the 3D flat metric, and \autoref{eq:scalarspecialconformal0} corresponds to choosing the constant vector $\vec c = (0, 0, 1)$. This is the large distance limit of the transformation effected by $K_3$. The corresponding transformation effected by $K_1$ or $K_2$ can be obtained by choosing $\vec c$ to point in the $x$ or $y$ direction. These transformations are precisely the standard special conformal transformations, for a field of conformal weight $1/2$, as is appropriate for a scalar in (effectively) three dimensions. The above transformation corresponds to choosing
\begin{equation}
\xi^i = - \vec x^2 c^i + 2 c^j x_j x^i
\end{equation}
in \autoref{deltaxiphi}, with indices contracted with the flat metric, and $d=3$.

The special conformal transformations are symmetries of 
$S_{\text{bulk}}$ but not of $S_{\text{Love}}$, i.e. the bulk Lagrangian's variation is a total derivative, but
the Love Lagrangian's variation is not:
\begin{equation}
\label{deltabulk}
\delta{\cal L}_{\rm bulk} = \partial_i \Lambda^i
\quad , \quad \Lambda^i = - \frac{1}{2} \left(\xi^i \partial_j \phi
  \partial^j \phi + \frac{1}{6} \phi^2 \partial_i \partial_j \xi^j
\right) \, ,
\end{equation}
\begin{equation}
\label{deltaLove}
\delta {\cal L}_{\rm Love} = \sum_{\ell=0}^\infty \frac{\lambda_{\ell}}{\ell!} \partial_{(i_1}
                  \dots \partial_{i_\ell)^T}\delta_\xi \phi \,\, \partial^{(i_1}
                  \dots \partial^{i_\ell)^T}\phi \,\, \delta^{(3)}\left(
                  {\vec x}
                  \right) \, .
\end{equation}
Detailed discussions of \autoref{deltabulk} can be found in \autoref{a:MCKVs4}. For \autoref{deltaLove}, it is tempting to declare certain terms vanish, simply because they involve products of the delta function with powers of $\vec x$. For instance, consider the $\ell=0$ term: one might naively declare $\lambda_0 \, \delta_\xi \phi \,\, \phi \, \delta^{(3)} (\vec x)$ to be vanishing, but keep in mind that the $\phi$ field profile is arbitrary, and could diverge at the origin. If one had naively declared $\delta{\cal L}_{\rm Love}$ to be vanishing for $\ell = 0$, it can be checked that the corresponding conservation of Noether current (see \autoref{MCKV}) would be inconsistent with the equation of motion \autoref{scalareomEFT}.

It is worth noting that while $\delta {\cal L}_{\rm Love}$ {\it is not} a total derivative for special conformal transformations, it {\it is} a total derivative for rotations. Hence, the Love number operators breaks the former but respects the latter. A discussion can be found in \autoref{breakingSCT}.\footnote{The special conformal symmetries are also broken by tadpole terms, or hair terms, such as ${\cal L}_{\rm hair} = \sum_\ell Q^{i_1 ... i_\ell} \,\, \partial_{(i_1}... \partial_{i_\ell)^T} \phi \,\, \delta^{(3)} (\vec x)$, where $Q^{i_1 ... i_\ell}$ represents multipole moments of the object. See \cite{Hui2022} for a discussion.}

Two comments are in order before we close this section. First, it's worth emphasizing that the special conformal symmetries apply strictly in the static limit. Thus, we do not expect them to protect black holes from quantum corrections generating a non-zero $\lambda_\ell$, i.e. even if the external tidal field is static, quantum loops can involve non-static field configurations. Second, the story spelled out above is in the large $r$ limit. One could work out subleading corrections suppressed by powers of $r_s/r$, by including the departure of the gravitational background from Minkowski.

\section{Static Spin 1 Perturbations}

\label{s:SchwarzschildVector}

We now seek to understand whether the vanishing of spin 1 Love numbers in the static limit arises from a similar symmetry argument to that of the spin 0 case. To this end, we consider a massless vector field in the Schwarzschild background given by the action
\begin{equation}
\label{eq:GenericPhotonAction}
S = -\frac{1}{4}\int g^{\mu\nu}g^{\rho\sigma}F_{\mu\rho}F_{\nu\sigma}\sqrt{-g}\dd[4]{x},
\end{equation}
where $F_{\mu\nu}$ is the Maxwell field strength tensor,
\begin{equation}
F_{\mu\nu}=\partial_{\mu}A_{\nu}-\partial_{\nu}A_{\mu},
\end{equation}
and we continue to use the Schwarzschild metric defined in \autoref{eq:SchwarzschildLine}.

\subsection{Symmetries of the Effective Metric}
\label{symmetriesSpin1}

We once again specialize to static perturbations and set all time derivatives to zero (i.e. $F_{it}=\partial_{i}A_{t}$). For spacetimes such as Schwarzschild, with $\partial_t g_{\mu\nu} = 0$ and $g_{ti} = 0$, the action decouples for electrostatic and magnetostatic perturbations and can be expressed as the sum of a scalar piece and a vector piece with respect to two potentially different effective 3D metrics $\hat g_E {}_{ij}$ and $\hat g_B {}_{ij}$:
\begin{equation}
\label{eq:EMAction}
S = \frac{1}{2}\int \dd{t} \dd[3]{x} \sqrt{\hat{g}_E} \hat{g}_E^{ij}\partial_{i}A_{t}\partial_{j}A_{t} -\frac{1}{4}\int \dd{t} \dd[3]{x} \sqrt{\hat{g}_B} \hat{g}_B^{ij}\hat{g}_B^{kl}F_{ik}F_{jl}.
\end{equation}
Once again, the effective metrics are not merely the same as
the spatial components of the 4D metric. They are related to the full 4D metric by
\begin{equation}
    \sqrt{\hat{g}_E} \hat{g}_E^{ij} = -g^{tt} g^{ij} \sqrt{-g} \qquad \text{and} \qquad \sqrt{\hat{g}_B} \hat{g}_B^{ij} \hat{g}_B^{kl} = g^{ij} g^{kl} \sqrt{-g}.
  \end{equation}
Solving these equations reveals that the effective metrics for the electrostatic and magnetostatic portions of the action are in fact the same, with $\hat{g}_{ij} = \hat{g}_{E\: ij} = \hat{g}_{B\: ij}$ being
\begin{equation}
    \hat{g}_{ij} = \frac{1}{-g_{tt}} g_{ij}.
\end{equation}
For a Schwarzschild background, $\hat{g}_{ij}$ is given by the line element
\begin{equation}
\label{eq:3DSchwarzschildVectorEffectiveMetric}
\dd{\hat{s}}^{2}=\frac{r^{4}}{\Delta^{2}}\left(\dd{r}^{2}+\Delta\left(\dd{\theta}^{2}+\sin^{2}\theta\dd{\varphi}^{2}\right)\right).
\end{equation}

The even parity degree of freedom, $A_{t}$, thus behaves as a scalar living in this effective 3D metric. We will see that there are geometric symmetries acting on $A_t$ as a result. For the odd parity degree of freedom that lives inside $F_{ij}$, there are different ways to isolate it. One more laborious way is to carefully examine the components of $A_i$, which we will describe in an inset at the end of this section; it will help make contact with our treatment in the spin 2 case. The other way is faster and leads efficiently to the desired result, which we describe here.

The definition of $F_{ij}$ in terms of the gauge field $A_i$ implies the familiar identity:
\begin{equation}
\nabla_i (\epsilon^{ijk} F_{jk}) = 0 \, ,
\end{equation}
where $\epsilon^{ijk}$ is the Levi-Civita tensor associated to the 3D effective metric $\hat{g}_{ij}$.\footnote{In other words, $\epsilon^{ijk} = \epsilon^{ijk}_{\rm flat} / \sqrt{\hat g}$, with $\epsilon^{ijk}_{\rm flat}$ being the standard Levi-Civita in flat space. Thus, $\nabla_i (\epsilon^{ijk} F_{jk}) = \partial_i (\sqrt{\hat g} \epsilon^{ijk} F_{ijk})/\sqrt{\hat g} = \partial_i (\epsilon^{ijk}_{\rm flat} F_{ijk})/\sqrt{\hat g} = 0$. A useful identity is $\epsilon^{ijk} \epsilon^{l} {}_{jk} = 2 \hat g^{il}$.} One could alternatively dispense with the gauge field $A_i$, and treat $F_{ij}$ as the dynamical field, with the proviso that the above condition be satisfied. This can be implemented by introducing a Lagrange multiplier $\chi$ into the (odd) action:\footnote{An alternative way to proceed is to insist $A_i$ as the dynamical field, in which case one can think of adding to $-F^2/4$ the following: $(\partial_i \chi) \epsilon^{ijk} F_{jk} /2$. Including the factor of $\sqrt{\hat g}$ in the integration measure, one sees that this new term gives rise to a total derivative, thus justifying its addition. Integrating by parts reproduces \autoref{eq:EMAuxiliary}.}  
\begin{equation}
\label{eq:EMAuxiliary}
S_{\rm odd} = \int \dd{t} \dd[3]{x} \sqrt{\hat{g}} \left(-\frac{1}{4} \hat{g}^{ij} \hat{g}^{kl} F_{ik} F_{jl} - \frac{1}{2} \chi \nabla_i\left(\epsilon^{ijk} F_{jk}\right) \right).
\end{equation}
Solving the equation of motion for $F_{ij}$ implies
\begin{equation}
\label{eq:dualization}
F_{ij} = \epsilon_{ij}^{\hphantom{ij}k}\partial_k \chi \, .
\end{equation}
Plugging this into \autoref{eq:EMAuxiliary} gives\footnote{It is important we did not simply plug \autoref{eq:dualization} into $S_{\rm odd} = \int (-F^2/4) \sqrt{\hat g} \, d^3x$, in which case we would have obtained \autoref{eq:EMAuxiliary2} with the wrong sign. This is reminiscent of writing an action for a particle in a system with spherical symmetry. If we wanted to enforce the constraint that $m r^2 \dot{\theta}^2 = L$, where $L$ is the angular momentum, it is important to include this as a constraint with a Lagrange multiplier rather than plugging it directly into the action, otherwise the radial equation of motion picks up an incorrect sign in the potential.}
\begin{equation}
\label{eq:EMAuxiliary2}
S_{\rm odd} = \frac{1}{2} \int \dd{t} \dd[3]{x} \hat{g}^{ij} \left(\partial_t A_t \partial_j A_t + \partial_i \chi \partial_j \chi\right).
\end{equation}
Essentially, we have introduced a scalar $\chi$, which is dual to the field strength in 3D, to isolate the odd degree of freedom.

Putting everything together, \autoref{eq:EMAction} and \autoref{eq:EMAuxiliary2} tell us:
\begin{equation}
\label{eq:StaticEMAction}
S = \frac{1}{2} \int \dd{t} \dd[3]{x} \sqrt{\hat{g}} \hat{g}^{ij}\left(\partial_i A_t \partial_j A_t + \partial_i \chi \partial_j \chi\right).
\end{equation}
An electromagnetic field in a Schwarzschild background is therefore decomposable into two scalars, one even-parity, $A_t$, and one odd-parity, $\chi$, living in an effective 3D metric given by \autoref{eq:3DSchwarzschildVectorEffectiveMetric}. This effective metric (let's refer to it as the \textit{vector effective metric}) is conformally related to the \textit{scalar effective metric}, \autoref{eq:3DSchwarzschildScalarEffectiveMetric}, by
\begin{equation}
\left( \hat{g}_{\text{vector}} \right)_{ij} = \frac{r^{4}}{\Delta^{2}}
\left( \hat{g}_{\text{scalar}} \right)_{ij}.
\end{equation}
From \Cref{symmetriesSpin0}, we already know the melodic CKVs of the scalar effective metric. In \Cref{a:MCKVs3}, we show that its conformal cousin, the vector effective metric, will have the same melodic CKVs {\it if} the conformal factor relating them is (up to a constant normalization) $\hat R_{\rm scalar} / \hat R_{\rm vector}$, i.e.
\begin{equation}
\left( \hat{g}_{\text{vector}} \right)_{ij} \propto (\hat R_{\rm
  scalar} / \hat R_{\rm vector} ) \times
\left( \hat{g}_{\text{scalar}} \right)_{ij} \,.
\end{equation}
It can be checked that this is indeed the case:
\begin{equation}
\hat{R}_{\text{vector}} = -\frac{3 r_s^2}{2r^4}, \qquad \hat{R}_{\text{scalar}} = \frac{r_s^2}{2\Delta^2}, \qquad \text{and} \qquad \frac{r^4}{\Delta^2} = -3\left(\frac{\hat{R}_{\text{scalar}}}{\hat{R}_{\text{vector}}}\right).
\end{equation}
Thus, the same SO(3,1) worth of geometric symmetries from the spin 0 case apply to the even and odd sectors of the spin 1 case. The same set of $\xi^i$'s (from $J$'s and $K$'s in \autoref{eq:AllMCKVs1} - \autoref{eq:AllMCKVs6}) can be used to generate symmetry transformations on $\delta_\xi A_t$ and $\delta_\xi \chi$, according to \autoref{deltaxiphi}. Keep in mind that the $\nabla_i \xi^i$ factor in the transformation will be slightly different, because of the difference between the vector and scalar effective metrics.

In addition to the SO(3,1) symmetries $\delta_\xi A_t$ and SO(3,1) symmetries $\delta_\xi \chi$, the action \autoref{eq:StaticEMAction} has the following SO(2) symmetry:
\begin{equation}
\begin{pmatrix} A_t \\ \chi \end{pmatrix} \rightarrow \begin{pmatrix} \cos\alpha & -\sin\alpha \\ \sin\alpha & \cos\alpha \end{pmatrix} \begin{pmatrix} A_t \\ \chi \end{pmatrix}. 
\end{equation}
where $\alpha$ is an arbitrary constant parameterizing the rotation. This encodes the electric-magnetic duality for spin 1 perturbations around a Schwarzschild black hole.

\begin{framed}
\noindent We close this section by giving an alternative derivation of \autoref{eq:EMAuxiliary2}. Similar reasoning will be used in the spin 2 case. Let us spell out the gauge field components:
\begin{equation}
A_\mu = \begin{pmatrix} A_t \\ A_r \\ A_a  \end{pmatrix}
\end{equation}
where $A_a$ are the angular components, with $a \in \left\{1, 2\right\}$. This can be further decomposed into a transverse and longitudinal piece on the 2-sphere:
\begin{equation}
A_a = D_a A_E + \epsilon_a^{\hphantom{a}b} D_b A_O,
\end{equation}
where the indices $a, b, c, \dots$ are raised and lowered using the metric on the 2-sphere, $\gamma_{ab}$. Meanwhile, $D_a$ will be the covariant derivative associated to the metric $\gamma_{ab}$. This means that $A_E$ is even under parity transformations while $A_O$ is odd. Adopting the gauge $A_E = 0$, the odd action can be written in terms of $A_r$ and $A_O$ (alternatively, doing so can be thought of as writing the odd action in terms of gauge invariant variables $A_r -\partial_r A_E$ and $A_O$):
\begin{subequations}
\label{SoddF}
\begin{align}
S_{\rm odd} &= -\frac{1}{4} \int \dd{t} \dd[3]{x} \sqrt{\hat{g}} \hat{g}^{ij} \hat{g}^{kl} F_{ik} F_{jl},\\
&
\begin{aligned}
&= -\int \dd{t} \dd{r} \dd{\theta} \dd{\varphi} \sqrt{\gamma} \frac{\Delta}{2r^2} D_a A_r D^a A_r \\
&\hspace{0.4cm} + \frac{1}{2}\int \dd{t} \dd{r} \dd{\theta} \dd{\varphi} \sqrt{\gamma} \left(\frac{\Delta}{r^2} \partial_r
  A_O \partial_r D^2 A_O - \left(D^2 A_O\right)^2\right).
\end{aligned}
\end{align}
\end{subequations}
The actions for $A_r$ and $A_O$ decouple. Indeed, the equation of motion for $A_r$ is simply 
\begin{equation}
D^2 A_r = 0. 
\end{equation}
This equation allows us to set $A_r = 0$, as long as the angular momentum $\ell\ne 0$. We have therefore isolated a single odd-parity degree of freedom, $A_O$, described by the action
\begin{equation}
\label{AOaction}
S_{\rm odd} = \frac{1}{2}\int \dd{t} \dd{r} \dd{\theta} \dd{\varphi} \sqrt{\gamma} \left(\frac{\Delta}{r^2} \partial_r A_O \partial_r D^2 A_O - \left(D^2 A_O\right)^2\right).
\end{equation}

This action contains up to four derivatives. To simplify it, we introduce an auxiliary field $\chi$:
\begin{equation}
S_{\rm odd} = \int \dd{t} \dd{r} \dd{\theta} \dd{\varphi} \sqrt{\gamma} \left[\chi \partial_r D^2 A_O + \frac{r^2}{2\Delta}
  D_a \chi D^a \chi - \frac{1}{2r^2} D^2 A_O D^2 A_O\right]. 
\end{equation}
The $\chi$ and $A_O$ equations of motion are
\begin{equation}
\label{eq:AOandchiRelationship}
\partial_r A_O = \frac{r^2}{\Delta} \chi \qquad \text{and} \qquad \partial_r \chi = -\frac{1}{r^2} D^2 A_O.
\end{equation}
If we use the $\chi$ equation of motion to integrate out $\chi$, we recover the original action for $A_O$ \autoref{AOaction}. If we instead integrate out $A_O$, we obtain an action for $\chi$, in which $\chi$ behaves as a scalar living in the background of the 3D effective metric described by the line element \autoref{eq:3DSchwarzschildVectorEffectiveMetric},
\begin{equation}
\label{eq:ActionforChi}
S_{\rm odd} = \frac{1}{2} \int \dd{t} \dd[3]{x} \sqrt{\hat{g}} \hat{g}^{ij} \nabla_i \chi \nabla_j
\chi.
\end{equation}
This thus describes a different route to \autoref{eq:EMAuxiliary2}.
\end{framed}

\subsection{Implications in Harmonic Space}

As in the spin 0 case, the representation theory of SO(3,1) tells us, among other things, solutions at different values of the angular momentum $\ell$ are connected by a ladder structure. We will start with the even-parity degree of freedom $A_t$. Making use of the same melodic CKV $K_3$ as before, and defining $\xi^i$ by \autoref{eq:MainMCKV}, we can work out the symmetry variation of $A_t$ following \autoref{deltaxiphi}:
\begin{equation}
\label{eq:VariationOfAt}
\delta_{{\xi}} A_t = \Delta \cos\theta \, \partial_r A_t + \frac{\Delta'}{2}\partial_\theta\left(\sin\theta A_t\right) - \frac{r_s}{2} \cos\theta A_t. 
\end{equation}
The symmetry variation of $A_t$ is equivalent to that of the spin 0 scalar $\phi$ up to the inclusion of an additional $\cos\theta$ term. Therefore, when $A_t$ is decomposed into spherical harmonics, the structure is the same as \autoref{eq:LadderfromK3},
\begin{equation}
\delta_{{\xi}} \left(A_t^{\left(\ell\right)} Y_{\ell}^{m}\right) = -f\left(\ell\right) D_\ell^+ A_t^{\left(\ell\right)} Y_{\ell + 1}^m + f\left(\ell - 1\right) D_{\ell}^- A_t^{\left(\ell\right)} Y_{\ell - 1}^{m},
\end{equation}
except the ladder operators $D_{\ell}^{\pm}$ are modified to be 
\begin{equation}
D_{\ell}^+ \equiv -\Delta \partial_r - \frac{\ell + 1}{2}\Delta' + \frac{r_s}{2} \qquad \text{and} \qquad D_{\ell}^- \equiv \Delta \partial_r - \frac{\ell}{2}\Delta' - \frac{r_s}{2}.
\end{equation}

The equation of motion can be written as:
\begin{equation}
H_{\ell} A_t^{\left(\ell\right)} = 0 \qquad \text{where} \qquad
H_{\ell} \equiv -\Delta\left(\frac{\Delta}{r^2} \partial_r \left(r^2
    \partial_r\right) - \ell\left(\ell + 1\right)\right) \, ,
\end{equation}
where $H_\ell$ can be thought of as a Hamiltonian operator. The intertwining relation is 
\begin{equation}
D_{\ell + 1}^- D_{\ell}^+ - D_{\ell - 1}^+ D_{\ell}^- = \frac{2\ell + 1}{4} r_s^2
\end{equation}
and so the Hamiltonian can be expressed in terms of the ladder operators as
\begin{equation}
H_{\ell} = D_{\ell + 1}^- D_{\ell}^+ + \frac{\ell\left(\ell + 2\right)}{4} r_s^2 = D_{\ell - 1}^+ D_{\ell}^- + \frac{\ell^2 - 1}{4} r_s^2, 
\end{equation}
which then leads to the same ladder algebra in \autoref{eq:LadderAlgebra}. This means $D^+_\ell$ and $D^-_\ell$ acting on a solution at level $\ell$ produces a solution at level $\ell+1$ and $\ell-1$ respectively.

At level $\ell = 1$, the Hamiltonian takes the factorized form $H_1 = D^+_0 D^-_1$. The ``ground state'' solution (at $\ell=1$ for spin 1, as opposed to $\ell=0$ for spin 0) is annihilated by the lowering operator:
\begin{equation}
\label{groundstateAt}
D_{1}^- A_{t}^{(1)} = 0,
\end{equation}
which is solved by
\begin{equation}
A_{t}^{(1)} \propto \frac{\Delta}{r}.
\end{equation}
This solution is regular at the horizon, which we select as a proper boundary condition since $A_t$ is gauge invariant in the static limit. Additionally, far away from the black hole, this mode is growing, with no decaying tail. As discussed in detail in the spin 0 case, the fact that the ground state solution satisfies a {\it first order} \autoref{groundstateAt} removes the puzzle of why it does not have both growing and decaying behavior at large $r$. By acting on the ground state solution with a successive raising operators, and by inspecting the form of the raising operators, it can be seen that the solutions at all $\ell \ge 1$ are regular at the horizon and purely growing at large $r$, with no decaying tail. It follows that the Love numbers vanish for $\ell \ge 1$.\footnote{An $\ell=0$ mode is not considered a tidal field for spin 1, just like an $\ell=0,1$ mode is not considered a tidal field for spin 2. The spin 1 solution derived using our approach agrees with the spin 1 even parity mode in \cite{Hui2021} who used a different field variable.}

Let us now turn our attention to the odd parity mode $\chi$, where the effective metric and, as a result, the symmetry variation are the same as before: 
\begin{equation}
\delta_{\xi} \chi = \Delta \cos\theta \partial_r \chi + \frac{\Delta'}{2}\partial_\theta\left(\sin\theta \chi\right) - \frac{r_s}{2} \cos\theta \chi.
\end{equation}
In addition, the decomposition into ladders and the associated Hamiltonian derived from the equation of motion are the same as for $A_t$. In order for the rest of the story to fully translate, we must make sure regularity at the horizon is also the appropriate boundary condition for $\chi$. Recall that $A_O$ is gauge invariant (see parenthetical remarks above \autoref{SoddF}), so we should insist on the regularity of $A_O$ at the horizon. Invoking the relation between $A_O$ and $\chi$ from \autoref{eq:AOandchiRelationship} thus tells us $\chi$ must be regular at the horizon as well. The rest of the argument for the vanishing of the Love numbers, for $\ell \ge 1$, carries through for the parity odd mode $\chi$.

It is worth highlighting that the symmetries and associated ladder structures we find here are able to be applied directly to the gauge fields. This is in contrast with the already known ladder symmetries of \cite{Hui2022}, where the ladders were written for the Teukolsky variable $\Phi^{(s)}$. It is possible, however, to relate the ladders found here to the ladders on the Teukolsky  variable. In the process of finding this connection, we found interesting identities for conformal Killing tensors, which we further discuss in \autoref{s:HigherSpinLadder}.

\subsection{Application to the Worldline EFT}

Let's turn our attention to the worldline EFT in the static limit. The initial setup is the same as in \Cref{eftspin0}. Following the discussion there, we will focus on the long distance limit in which the geometry can be approximated as flat. The relevant action, including a bulk piece and a worldline piece for the spin 1 field, is:
\begin{equation}
S = S_{\text{bulk}} + S_{\text{Love}} \, ,
\end{equation}
where
\begin{equation}
S_{\text{bulk}} = \int \left(\frac{1}{2}\partial_i A_t \partial^i A_t - \frac{1}{4} F_{ij} F^{ij} - \frac{1}{2} \chi \partial_i (\epsilon^{ijk} F_{jk}) \right) \dd[4]{x},
\end{equation}
and 
\begin{equation}
\begin{aligned}
S_{\text{Love}} &= \sum_{\ell = 1}^{\infty}m\frac{\lambda^{(E)}_{\ell}}{2\ell!} \int \partial_{(i_1} \dots \partial_{i_{\ell - 1}} E_{i_\ell)^T} \partial^{(i_1} \dots \partial^{i_{\ell - 1}} E^{i_\ell)^T} \delta^{(3)}\left(\vec x\right) \dd[4]{x}\\
&\hspace{0.5cm} + \sum_{\ell = 1}^{\infty} \frac{\lambda^{(B)}_{\ell}}{4\ell!} \int \partial_{(i_1} \dots \partial_{i_{\ell - 1}} F_{i_{\ell} j)^T} \partial^{(i_1} \dots \partial^{i_{\ell - 1}} F^{i_{\ell} j)^T} \delta^{(3)}\left(\vec x\right) \dd[4]{x}.
\end{aligned}
\end{equation}
Here, the point object of interest is taken to be at rest at the origin. All time derivatives are set to zero. The bulk action comes from the flat-space limit of \autoref{eq:EMAction} and \autoref{eq:EMAuxiliary}. The Love number terms in $S_{\rm Love}$ encapsulate the finite-size effect of tidal deformability, or in electromagnetism, more often referred to as polarizability. There are two kinds: one for electric (parity even) polarizability, and the other for magnetic (parity odd) polarizability. The electric and magnetic Love numbers (polarizabilities) are labeled as $\lambda_{\ell}^{(E)}$ and $\lambda_{\ell}^{(B)}$.

The next step is to integrate out $F_{ij}$ as we did in \Cref{symmetriesSpin1}. The main difference from before is the presence of the Love number terms. Solving the $F_{ij}$ equation of motion for $F_{ij}$ in terms of $\chi$ --- perturbatively in $\lambda_\ell^{(B)}$ ---it can be shown that to first order in polarizability the action takes the expected form:
\begin{equation}
\label{eq:EMcombined}
\begin{aligned}
S & = \frac{1}{2} \int \delta^{ij}
\left(\partial_{i}A_{t}\partial_{j}A_{t} +
  \partial_{i}\chi\partial_{j}\chi\right) \dd[4]{x} \, \\
& \hspace{0.5cm} + \sum_{\ell = 1}^{\infty}
\frac{\lambda_{\ell}^{(E)}}{2\ell!} \int \partial_{(i_1} \dots
\partial_{i_{\ell})^T} A_t \,\, \partial^{(i_1} \dots
\partial^{i_{\ell})^T} A_t \,\, \delta^{(3)}\left(\vec x\right)
\dd[4]{x} \\
&\hspace{0.5cm} + \sum_{\ell = 1}^{\infty}
\frac{\lambda_{\ell}^{(B)}}{2\ell!} \int \partial_{(i_1} \dots
\partial_{i_{\ell})^T} \chi \,\, \partial^{(i_1} \dots
\partial^{i_{\ell})^T} \chi \,\, \delta^{(3)}\left(\vec x\right)\dd[4]{x}
\end{aligned}
\end{equation}

Just as in the spin 0 case, solving the $A_t$ or $\chi$ equation of motion, and matching the result against the UV computation for a black hole, tells us that the spin 1 Love numbers for a black hole are zero. This raises the question: what symmetry could forbid the existence of Love number operators in the EFT described by \autoref{eq:EMcombined}\,? Or: if $\lambda_\ell^{(E)}$ or $\lambda_\ell^{(B)}$ were non-zero, what symmetry is broken?

And just as in the spin 0 case, we take a cue from the UV computation for the black hole. Taking the large $r$ limit of \autoref{eq:VariationOfAt}, the symmetry variation of $A_t$ generated by $K_3$ is
\begin{equation}
    \delta_\xi A_t \rightarrow r^2 \cos\theta \, \partial_r A_t + r\partial_\theta\left(\sin\theta A_t\right),
  \end{equation}
and similarly so for $\chi$. This is simply a special conformal transformation on $A_t$ or $\chi$ (as scalar), written in spherical coordinates. The full set of special conformal symmetries (generated by the melodic CKVs $K_1$, $K_2$, $K_3$) is most simply expressed in Cartesian coordinates: simply replace $\phi$ in \autoref{eq:scalarspecialconformal2} by $A_t$ or $\chi$. The special conformal transformations are symmetries of the bulk term in \autoref{eq:EMcombined} but not of the Love number terms, following the same reasoning as in the spin 0 case.

It's worth stressing that $\chi$, the dual scalar we introduced via \autoref{eq:dualization}, is related to the gauge field in a non-local way (see also \autoref{eq:AOandchiRelationship}). Thus, the special conformal symmetry on $\chi$ translates into a non-local symmetry on the gauge field.

\section{Static Spin 2 Perturbations}
\label{s:Spin2}

In this section, we aim to apply the same symmetry arguments to explain the vanishing of the spin 2 Love numbers in the static limit. The action is given by perturbing around a background metric, $\bar{g}_{\mu\nu}$, in the Einstein--Hilbert action:
\begin{equation}
\label{eq:GenericGravitonAction}
S = \int R\left[\bar{g}_{\mu\nu} + \delta g_{\mu\nu}\right]\sqrt{-\left(\bar{g} + \delta g\right)}\dd[4]{x}.
\end{equation}
The background metric in this case will still be the Schwarzschild metric, \autoref{eq:SchwarzschildLine}. 

\subsection{Symmetries of the Effective Metrics}
\label{Symm_spin2}
In the spin 1 case, we were able to rewrite the action in terms of two scalars, one odd parity and one even parity, each living in an effective 3D background possessing melodic conformal Killing vectors. In the process of doing so, we found that the theory of a 4D vector field in four dimensions decouples in the static case into a 3D vector field and a scalar field. Following the same line of thinking, we'd like to decompose the full tensor action into the action for a tensor field, a vector field, and scalar field in one lower dimension. This is done efficiently using the \textit{Kaluza--Klein decomposition}, which we review in \autoref{a:KKDecomp}. We again expect to find two physical degrees of freedom: one even parity mode and one odd parity one. Indeed, in this subsection, we will identify the two physical degrees of freedom, one even-parity mode, $\delta \phi$, and one odd-parity mode, $\chi$, resulting in the action for these two scalars given by \autoref{eq:FinalEvenOddSpin2Action}, with the two effective metrics given by \autoref{eq:3DSchwarzschildOddTensorEffectiveMetric} and \autoref{eq:3DSchwarzschildEvenTensorEffectiveMetric} respectively. We will also show that the actions for the even and odd parity scalars each have the same SO(1, 3) symmetry as in the spin 0 and spin 1 case with an additional SO(2) symmetry that relates the two sectors. 

As per \autoref{eq:KKAction}, the Kaluza--Klein decomposition of the Einstein-Hilbert action into a tensor, vector, and scalar piece in one lower dimension is
\begin{equation}
    S = \int \dd{t} \dd[3]{x} \sqrt{g_3} \left(R_3 - \frac{1}{2}\partial_i \phi \partial^i \phi + \frac{1}{4} e^{-2\phi} F^{ij} F_{ij}\right),
\end{equation}
where the four-dimensional metric is parameterized as in \autoref{eq:KKMetric},
\begin{equation}
    \dd{s}^2 = -e^{-\phi}\left(\dd{t} + A_i \dd{x}^i\right)^2 + e^\phi g_{3\:ij} \dd{x}^i \dd{x}^j,
\end{equation}
and $F_{ij} = \partial_i A_j - \partial_j A_i$. We will refer to the collection of $\phi$, $A_i$, and $g_{3\: ij}$ as the Kaluza--Klein variables. Our next step will be to obtain the actual modes of interest by expanding this about the Schwarzschild background to second order. In relating this metric to the Schwarzschild metric, we find that the background values for the Kaluza--Klein variables are
\begin{equation}
\label{eq:BackgroundEqns}
    e^{\bar{\phi}} = \frac{r^2}{\Delta}, \qquad \bar{A}_i = 0, \qquad \text{and} \qquad \dd{\bar{s}}^2_3 = \dd{r}^2 + \Delta \dd{\Omega}^2,
\end{equation}
where the backgrounds are indicated by a bar. The perturbations will then be $\delta\phi$, $A_i$ (since the background is zero), and $\delta g_{3\: ij} = h_{ij}$. If we expand the action to second order, we find that it decomposes into a scalar, vector, tensor, and interaction piece,
\begin{equation}
    S = S_{\text{scalar}} + S_{\text{vector}} + S_{\text{tensor}} + S_{\text{int}}.
\end{equation}
Let's begin with the vector piece, which is already at second order in perturbations since the background $\bar{F}_{ij} = 0$:
\begin{equation}
    S_{\text{vector}} = \frac{1}{4}\int \dd{t} \dd[3]{x} \sqrt{\bar{g}_3} e^{-2\bar{\phi}} \bar{g}_3^{ij} \bar{g}_3^{kl} F_{ik} F_{jl}.
\end{equation}
This completely decouples from the scalar and tensor pieces. The remaining three pieces are:
\begin{subequations}
\begin{align}
    S_{\text{scalar}} &= -\frac{1}{2} \int \dd{t} \dd[3]{x} \sqrt{\bar{g}_3} \bar{g}^{ij}_3 \: \partial_i \delta \phi \partial_j \delta \phi,\\
    \label{eq:TensorAction}
    S_{\text{tensor}} &= \frac{1}{4} \int \dd{t} \dd[3]{x} \sqrt{\bar{g}_3} \left(\nabla_i h \nabla^i h - 2 \nabla_j h \nabla_i h^{ij} + 2\nabla_k h_{ij} \nabla^j h^{ik} - \nabla_k h_{ij} \nabla^k h^{ij}\right), \\
    S_{\text{int}} &= -\int \dd{t} \dd[3]{x}\sqrt{\bar{g}_3} \left(h^{ij} - \frac{1}{2} h \bar{g}^{ij}_3\right) \partial_i \delta \phi \partial_j \bar{\phi}. \label{eq: spin 2 mixing}
\end{align}
\end{subequations}
Here, all covariant derivatives are those of the background metric $\bar{g}_{3\:ij}$. In simplifying the tensor portion of the action in particular, it is helpful to observe that
\begin{equation}
    \bar{R}_{3\: ij} = \frac{1}{2}\partial_i \bar{\phi} \: \partial_j \bar{\phi}.
\end{equation}

Let's first study the vector action and find out whether that can be written in terms of a scalar living in an effective metric. This proceeds in analogy to the spin 1 case, where we express $F_{ij}$ in terms of the odd-parity field, $\chi$, as in \autoref{eq:dualization} and rescale the metric $g_{3\: ij}$ to write the action in terms of the scalar $\chi$ living in a different 3D background, 
\begin{equation}
    S_{\text{vector}} = -\frac{1}{2} \int \dd{t} \dd[3]{x} \sqrt{\hat{g}_O} \hat{g}_{\text{O}}^{ij} \partial_i \chi \partial_j \chi.
\end{equation}
The subscript $O$ indicates that this is the effective metric for the odd-parity metric perturbation. The line element associated with this metric is
\begin{equation}
\label{eq:3DSchwarzschildOddTensorEffectiveMetric}
\dd{\hat{s}}^2_O = \frac{r^8}{\Delta^4}\left(\dd{r}^2 + \Delta \left(\dd{\theta}^2 + \sin^2\theta \dd{\varphi}^2\right)\right).
\end{equation}

Having isolated the odd parity degree of freedom as a minimally coupled scalar living in an effective 3D metric we can turn our attention to the tensor, scalar, and interaction parts of the action. By the electric-magnetic duality of the Schwarzschild background, we expect that the remaining pieces of the action should reduce to an even-parity scalar living in a potentially different effective metric \cite{Chandrasekhar:1985}. This motivates us to decompose the tensor pieces as done in \autoref{a:GIVariables}. First, the tensor piece $h_{ij}$ can be decomposed into $h_{rr}$, $h_{ra}$, and $h_{ab}$ where $a$, $b$, $c$, etc. are angular indices that are raised and lowered with the 2-sphere metric, $\gamma_{ab}$ given by:
\begin{equation}
    \dd{s}^2_\gamma = \dd{\theta}^2 + \sin^2\theta \dd{\varphi}^2.
\end{equation}
Next, $h_{ra}$ can be further decomposed into an even and odd parity piece while $h_{ab}$ can be decomposed into a trace, trace-free even, and trace-free odd piece, defined as follows:
\begin{equation}
\begin{aligned}
    h_{ra} &= D_{a} \mathcal{H}_E + \epsilon_{a}^{\hphantom{a}b} \mathcal{H}_O,\\
    h_{ab} &= \frac{\Delta}{2}\gamma_{ab}\mathcal{K} + \left(D_{a} D_{b} - \frac{1}{2}\gamma_{ab} D^2\right)\mathcal{G}_E + \frac{1}{2}\left(\epsilon_{a}^{\hphantom{a}c} D_{c} D_{b} + \epsilon_{b}^{\hphantom{b}c}D_{c} D_{a}\right)\mathcal{G}_O,
\end{aligned}
\end{equation}
where $D_{a}$ is the derivative on the 2-sphere.

We now consider combinations of these variables that are independent of gauge, aptly named gauge-independent variables \citep{Martel2005}. The action and equations of motion, being gauge-invariant, can then be recast entirely in terms of these variables. This simplifies the calculation and helps us to isolate the even parity scalar living in an effective metric while avoiding subtleties related to whether a gauge choice in the action is complete \citep{Motohashi2016}. Since we have already rewritten the vector action in terms of a scalar, we will solely focus on the terms within $h_{ij}$ and $\delta\phi$. The gauge independent variables from \autoref{eq:OddGIV} and \autoref{eq:EvenGIV} are
\begin{subequations}
\begin{align}
\label{eq:tildephi}
    \delta\tilde{\phi} &= \delta \phi - \frac{r_s}{\Delta}\mathcal{H}_E + \frac{r_s}{2}\partial_r \left(\frac{1}{\Delta}\mathcal{G}_E\right),\\
    \tilde{\mathcal{K}} &= \mathcal{K} - \frac{2\Delta'}{\Delta}\mathcal{H}_E + \Delta' \partial_r\left(\frac{1}{\Delta}\mathcal{G}_E\right) - \frac{1}{\Delta}D^2 \mathcal{G}_E,\\
    \tilde{h}_{rr} &= h_{rr} - 2\partial_r\mathcal{H}_E + \partial_r\left(\Delta \partial_r\left(\frac{1}{\Delta}\mathcal{G}_E\right)\right),\\
    \tilde{\mathcal{H}}_O &= \mathcal{H}_O - \frac{\Delta}{2}\partial_r\left(\frac{1}{\Delta}\mathcal{G}_O\right).
    \label{eq:tildeHO}
\end{align}
\end{subequations}

Our goal is to recast the action in terms of these variables. A shortcut to achieving this is to exploit the gauge-invariant nature of these variables by making any convenient gauge choice that simplifies the right-hand side of \autoref{eq:tildephi}-\autoref{eq:tildeHO} and reduces the number of  terms in the action. From looking at the gauge transformations in \autoref{eq:FinalGaugeTrans}, we see that a particularly nice gauge choice is 
\begin{equation}
\label{eq:RWGauge}
    \mathcal{H}_E = \mathcal{G}_E = \mathcal{G}_O = 0.
\end{equation} 
This is the same as the original Regge--Wheeler gauge \cite{Regge:1957}, though other popular gauge choices exist, such as the Regge--Wheeler-unitary gauge \cite{Franciolini:2018}. Note that we have only used three gauge degrees of freedom. The fourth one, arising from time diffeomorphisms, $\xi_t$, lives entirely in the vector sector and corresponds to the U(1) gauge symmetry of the vector field $A_{i}$, as shown in \autoref{a:GIVariables}. The advantage of picking this gauge is that the gauge independent variables become very simple:
\begin{equation}
    \delta\tilde{\phi} = \delta \phi, \quad \tilde{\mathcal{K}} = \mathcal{K}, \quad \tilde{h}_{rr} = h_{rr}, \quad \text{and} \quad \tilde{\mathcal{H}}_O = \mathcal{H}_O.
\end{equation}
At any point, if one wants to recover a gauge-independent action or equations of motion, all that needs to be done is insert a tilde on top of these remaining variables.

Let's begin by focusing on the \textit{even sector} of the tensor, scalar, and interaction actions by fixing this gauge and ignoring terms that arise from $\mathcal{H}_O$, as those will decouple. Explicitly, this means utilizing the following perturbation tensor
\begin{equation}
    h_{ij, \: \text{even}} = \begin{pmatrix}
    h_{rr} & 0 & 0 \\ 
    0 & \displaystyle{\frac{\Delta}{2}\mathcal{K}} & 0 \\
    0 & 0 & \displaystyle{\frac{\Delta}{2}\mathcal{K} \sin^2\theta}
    \end{pmatrix}.
\end{equation}
Plugging this into the scalar, tensor, and interaction actions and simplifying by using integration by parts yields:
\begin{equation}
\begin{aligned}
    S = -\frac{1}{2}\int \dd{t} \dd[3]{x} \sqrt{\gamma} &\Big[\Delta\left(\partial_r\delta\phi\right)^2 - \frac{\Delta}{4}\left(\partial_r\mathcal{K}\right)^2 - r_s \delta\phi\partial_r \mathcal{K} - r_s h_{rr}\partial_r\delta\phi \\
    & - \frac{\Delta'}{2}\mathcal{K}\partial_rh_{rr} - h_{rr}^2 - \frac{1}{2}D_{a}h_{rr}D^{a}\mathcal{K} + D_{a}\delta\phi D^{a}\delta \phi\Big].
\end{aligned}
\end{equation}
In order to further simplify this action, we will integrate out the extra degrees of freedom by using the equations of motion. The constraint equations in this static case will correspond to equations that are less than quadratic in the radial derivative. The $h_{rr}$, $K$, and $\delta \phi$ equations of motion respectively are:
\begin{subequations}
\begin{align}
        -r_s\partial_r\delta \phi - 2h_{rr} + \frac{1}{2}\partial_r\left(\Delta' \mathcal{K}\right) + \frac{1}{2} D^2 \mathcal{K} &= 0,\\
        \partial_r\left(r_s \delta \phi + \frac{\Delta}{2}\partial_r\mathcal{K}\right) - \frac{\Delta'}{2}\partial_r h_{rr} + \frac{1}{2}D^2 h_{rr} &= 0,\\
        \label{eq:TensorEoM}
        \partial_r\left(\Delta \partial_r\delta \phi\right) + \frac{r_s}{2}\partial_r\mathcal{K} - \frac{r_s}{2}\partial_r h_{rr} + D^2 \delta \phi &= 0.
\end{align}
\end{subequations}
These three equations can appropriately be combined into the following constraint equation:
\begin{equation}
    D^2\left(r_s \delta \phi + \frac{\Delta}{2} \partial_r\mathcal{K} - \frac{\Delta'}{2} h_{rr}\right) = 0.
\end{equation}
Since the Laplacian of this function on the sphere is $0$, the function itself must also be $0$, so
\begin{equation}
\label{eq:ConstraintPhiK}
    r_s \delta\phi = \frac{\Delta'}{2}h_{rr} - \frac{\Delta}{2}\partial_r\mathcal{K}.
\end{equation}

If we plug this constraint into the second equation, we find the following equation for $h_{rr}$:
\begin{equation}
    \left(D^2 + 2\right)h_{rr} = 0.
\end{equation}
There are two solutions to this equation. Either $h_{rr}$ is 0, or there is a fixed dipole solution with $\ell = 1$. Indeed, the dipole solution corresponds to a perturbation that shifts the position of the black hole \cite{Thorne2019}, but since we are only interested in \textit{dynamical} degrees of freedom we can, without loss of generality, set $h_{rr}$ to $0$. The second constraint relates $\partial_r\mathcal{K}$ with $\delta \phi$. Writing everything in terms of $\delta \phi$, we find
\begin{equation}
    S = -\frac{1}{2}\int \dd{t} \dd[3]{x} \sqrt{\hat{g}_E} \left(\hat{g}_E^{ij} \partial_i \delta \phi \partial_j \delta \phi + 2\hat{R}_E \delta\phi^2\right),
\end{equation}
where $\hat{g}_{E\: ij}$ is exactly the same as the static scalar effective metric, given by the line element
\begin{equation}
\label{eq:3DSchwarzschildEvenTensorEffectiveMetric}
    \dd{\hat{s}}_E^2 = \dd{r}^2 + \Delta \left(\dd{\theta}^2 + \sin^2\theta\dd{\varphi}^2\right),
\end{equation}
and $\hat{R}_E$ is the Ricci scalar associated to this metric, given in \autoref{eq:StaticScalarRicci}. The presence of this non-minimal coupling term does not change the number of symmetries generated by melodic CKVs that the action has. As reviewed in \Cref{a:MCKVs1}, the melodic CKVs are symmetries for arbitrary coupling to the Ricci scalar, including minimal coupling and non-conformal couplings.

Before turning our attention to the melodic CKVs associated with these two metrics, we should study the odd sector of the tensor and scalar actions. Utilizing the fact that $\epsilon_\theta^{\hphantom{\theta}\varphi} = 1/\sin\theta$ and $\epsilon_\varphi^{\hphantom{\varphi}\theta} = -\sin\theta$, the odd sector perturbation metric is
\begin{equation}
    h_{ij, \: \text{odd}} = \begin{pmatrix} 0 & \displaystyle{\frac{1}{\sin\theta}\partial_\varphi \mathcal{H}_O} & \displaystyle{-\sin\theta \, \partial_\theta \mathcal{H}_O} \\ \displaystyle{\frac{1}{\sin\theta}\partial_\varphi \mathcal{H}_O} & 0 & 0 \\
    \displaystyle{-\sin\theta \,\partial_\theta \mathcal{H}_O} & 0 & 0 \end{pmatrix}.
\end{equation}
The action in this case, upon integration by parts, reduces to
\begin{equation}
    S = \frac{1}{2}\int\frac{1}{\Delta}\left(\left(D^2 \mathcal{H}_O\right)^2 - 2D_{a} \mathcal{H}_O D^{a} \mathcal{H}_O\right)\sqrt{\gamma}\dd[3]{x}.
\end{equation}
The equation of motion in this case is simply
\begin{equation}
    D^2\left[\left(D^2 + 2\right)\mathcal{H}_O\right] = 0.
\end{equation}
One solution to this equation is an $\ell = 1$ mode, which is not considered to be a tidal field for spin 2. Instead, this corresponds to a perturbation in the spin of the black hole \citep{Thorne2019}. Since we are only interested in the modes corresponding to tidal fields, we can safely set this to $0$.

At this point, we've shown that the dynamics of static tensor perturbations around a Schwarzschild black hole can be entirely captured by two scalars, $\chi$ and $\delta\phi$ living in the effective metrics \autoref{eq:3DSchwarzschildOddTensorEffectiveMetric} and \autoref{eq:3DSchwarzschildEvenTensorEffectiveMetric} respectively. The total action describing the static spin 2 modes is given by
\begin{equation}
\label{eq:FinalEvenOddSpin2Action}
S = -\frac{1}{2}\int \dd{t} \dd[3]{x} \sqrt{\hat{g}_E} \left(\hat{g}_E^{ij} \partial_i \delta\phi \partial_j \delta \phi + 2\hat{R}_E \delta\phi^2\right) - \frac{1}{2} \int \dd{t} \dd[3]{x} \sqrt{\hat{g}_O} \hat{g}_O^{ij} \partial_i \chi \partial_j \chi.
\end{equation}

As was the situation in the spin 1 case, the effective metric for the odd parity mode is conformally related to the static scalar effective metric. Therefore, if the conformal factor is proportional to the ratio of the Ricci Scalar $\hat{R}_{\text{scalar}}$ and the Ricci scalar of the spin 2 odd effective metric, the melodic CKVs of the new effective metrics will be the same as the melodic CKVs of the scalar effective metric, \autoref{eq:AllMCKVs1} - \autoref{eq:AllMCKVs6}. In particular, the Ricci scalar of the odd effective metric, Ricci scalar of the scalar effective metric are, and the relationship between the conformal factor relating the even and off metrics is:
\begin{equation}
\label{R0Rscalar}
    \hat{R}_O = -\frac{15r_s^2 \Delta^2}{2r^8}, \qquad \hat{R}_{\text{scalar}} = \frac{r_s^2}{2\Delta^2}, \qquad \text{and} \qquad \frac{r^8}{\Delta^4} = -15 \frac{\hat{R}_{\text{scalar}}}{\hat{R}_O}.
\end{equation}
Indeed, the last equation shows that the conformal factor is related to the ratio by a constant. The even effective metric is precisely the same as the static scalar effective metric, so it also has the melodic CKVs given by \autoref{eq:AllMCKVs1} - \autoref{eq:AllMCKVs6}.

To summarize, the geometric symmetry transformations on $\delta\phi$ and on $\chi$ are both given by \autoref{deltaxiphi}, with $\xi$ being one of the melodic CKVs. Note that the effective 3D metrics for $\delta\phi$ and $\chi$ are different (\autoref{eq:3DSchwarzschildOddTensorEffectiveMetric} and
\autoref{eq:3DSchwarzschildEvenTensorEffectiveMetric}), and that should be kept in mind when computing $\nabla_i \xi^i$. 

Another way to express the action is to explicitly include the Weyl factor,
\begin{equation}
\hat{g}_{ij} \equiv \hat{g}_{E\:ij} = \Omega^{-2} \hat{g}_{O\:ij},
\end{equation}
where $\Omega = r^4/\Delta^2$, so 
\begin{equation}
\label{eq:CombinedSpin2Action}
S = -\frac{1}{2} \int \dd{t}\dd[3]{x} \sqrt{\hat{g}} \left(\hat{g}^{ij} \partial_i \delta \phi \partial_j \delta \phi + 2\hat{R}_E \phi^2 + \Omega \hat{g}^{ij} \partial_i \chi \partial_j \chi\right).
\end{equation}
Written in this way, we see that in addition to the SO(3,1) symmetries the even and odd sectors have separately, there is an additional SO(2) symmetry:
\begin{equation}
\begin{pmatrix} 
\delta \phi \\ \chi 
\end{pmatrix} \rightarrow \begin{pmatrix}
\cos\alpha & -\Omega^{1/2} \sin\alpha \\
\Omega^{1/2} \sin\alpha & \cos\alpha 
\end{pmatrix} \begin{pmatrix} 
\delta \phi \\ \chi 
\end{pmatrix}.
\end{equation}
This encodes the Chandrasekhar duality that maps between even-parity and odd-parity black hole perturbations \cite{Solomon:2023}. A demonstration of this symmetry and derivation of the associated Noether current is contained in \autoref{a:Chandrasekhar}.

\subsection{Implications in Harmonic Space}
Since these effective metrics have the same SO(3,1) symmetry as in the previous two cases, we can construct a ladder that connects solutions at different values of $\ell$. The variation of the scalars $\delta\phi$ and $\chi$ under the melodic CKV \autoref{eq:MainMCKV} are
\begin{subequations}
\begin{align}
\delta_{\xi} \delta \phi &= \Delta \cos\theta \, \partial_r \delta \phi + \frac{\Delta'}{2}\partial_\theta\left(\sin\theta \, \delta \phi\right),\\
\delta_{\xi} \chi &= \Delta \cos\theta \, \partial_r \chi + \frac{\Delta'}{2}\partial_\theta\left(\sin\theta \, \chi \right) - r_s \cos\theta \, \chi.
\end{align}
\end{subequations}
Since $\delta\phi$ in this gauge is equivalent to the gauge-independent $\delta\tilde{\phi}$ (see the discussion above \autoref{eq:tildephi}), we will impose regularity on at the horizon for $\delta \phi$. For $\chi$, the story proceeds as in the previous section, namely that since $\chi$ is gauge-invariant we should also impose regularity on $\chi$ at the horizon.

Since the effective metric for $\delta \phi$ is the same as in the spin 0 case, the associated ladders are also the same as in \autoref{eq:Spin0Ladders}. The Hamiltonian, however, is different since the action contains a non-minimal coupling term. Namely, the radial equation of motion for $\delta\phi\left(r\right)$ at level $\ell$ is
\begin{equation}
H_{\ell} \delta \phi_{\ell} = 0, \qquad \text{where} \qquad H_{\ell} = -\Delta\left(\partial_r\left(\Delta \partial_r\right) - \ell \left(\ell + 1\right) - \frac{r_s^2}{\Delta}\right),
\end{equation}
which implies that the relationship between the Hamiltonian and the ladder operators is given by
\begin{equation}
H_{\ell} = D_{\ell + 1}^- D_{\ell}^+ + \frac{\left(\ell + 3\right)\left(\ell - 1\right)}{4} r_s^2 = D_{\ell - 1}^+ D_{\ell}^- + \frac{\ell^2 - 4}{4} r_s^2.
\end{equation}
Therefore, although the Hamiltonians are different, the ladders still retain the interpretation that $D_{\ell}^+$ and $D_{\ell}^-$ acting on a solution at level $\ell$ produces a solution at level $\ell + 1$ and $\ell - 1$ respectively. 

At level $\ell = 2$, the Hamiltonian factorizes, $H_2 = D_1^+ D_2^-$ and the ground state solution (now at $\ell = 2$ for spin 2 as opposed to $\ell = 1$ for spin 1 and $\ell = 0$ for spin 0) is annihilated by the lowering operator:
\begin{equation}
D_2^- \delta \phi_2 = 0,
\end{equation}
which is solved by
\begin{equation}
    \delta \phi_2 \propto \Delta.
\end{equation}
This ground state solution to a first-order differential equation is regular at the horizon and purely growing far away from the black hole. The ladder can then be used to construct the solutions at any level $\ell \geq 2$ and we are now guaranteed that they are the regular solution at the horizon and purely growing at infinity, meaning they have no decaying tail. Hence, the Love numbers vanish for all $\ell \geq 2$.\footnote{Recall again that the $\ell = 0$ and $\ell = 1$ modes are not considered tidal fields for spin 2. The $\ell = 0$ mode is considered a mass perturbation while the $\ell = 1$ mode is considered a spin perturbation, as discussed in \cite{Thorne2019}.} 

On the other hand, $\chi$ lives in a different effective metric. By decomposing $\chi$ into spherical harmonics, we find that the variation under $\xi$ implies the existence of the following ladders 
\begin{equation}
D_{\ell}^+ \equiv -\Delta \partial_r - \frac{\ell + 1}{2}\Delta' + r_s \qquad \text{and} \qquad D_{\ell}^- \equiv \Delta \partial_r - \frac{\ell}{2} \Delta' - r_s. 
\end{equation}
In this case, the equation of motion is given by
\begin{equation}
H_{\ell} \chi_{\ell} = 0 \qquad \text{where} \qquad H_{\ell} = -\Delta\left(\frac{\Delta^2}{r^4} \partial_r\left(\frac{r^4}{\Delta}\partial_r\right) - \ell\left(\ell + 1\right)\right).
\end{equation}
The intertwining relation is the same as \autoref{eq:MainIntertwiningRelation}, which implies that the Hamiltonian can be expressed as
\begin{equation}
H_{\ell} = D_{\ell + 1}^- D_{\ell}^+ + \frac{\left(\ell + 3\right)\left(\ell - 1\right)}{4} r_s^2 = D_{\ell - 1}^+ D_{\ell}^- + \frac{\ell^2 - 4}{4} r_s^2.
\end{equation}
Once again the Hamiltonian factorizes at level $\ell = 2$, where it simplifies into the first order equation with $\chi_2$ being annihilated by the lowering operator: 
\begin{equation}
D_2^- \chi_{2} = 0 \Rightarrow \chi_2 \propto \left(\frac{\Delta}{r}\right)^2. 
\end{equation}
Since the solution is regular at the horizon and purely growing at infinity, the ladders can be used to raise $\chi_2$ to any solution $\chi_{\ell}$ and we are guaranteed to find the solution that is regular at the horizon is also purely growing as $r^{\ell}$ at infinity. Since there is no decaying mode, the Love numbers vanish for all $\ell \geq 2$.

\subsection{Application to the Worldline EFT}
Having discussed the UV symmetries, we would like to study their application to the worldline EFT. Working in the approximation where the background is flat, the action can be decomposed into two pieces 
\begin{equation}
S = S_{\text{bulk}} + S_{\text{Love}},
\end{equation}
where 
\begin{equation}
\begin{aligned}
S_{\text{bulk}} = \int \dd{t} \dd[3]{x} &\Big(\frac{1}{4} \partial_i h \partial^i h - \frac{1}{2}\partial_j h \partial_i h^{ij} + \frac{1}{2}\partial_k h_{ij} \partial^j h^{ik} \\
&\hspace{0.2cm} - \frac{1}{4}\partial_k h_{ij} \partial^k h^{ij} - \frac{1}{2} \partial_i \delta \phi \delta^i \delta \phi + \frac{1}{4} F^{ij} F_{ij} \Big). 
\end{aligned}
\end{equation}
and 
\begin{equation}
\begin{aligned}
S_{\text{Love}} &= \sum_{\ell = 2}^{\infty} \frac{\lambda_{\ell}^{(C_E)}}{2\ell!} \int \dd{t} \dd[3]{x} \partial_{(i_1} \dots \partial_{i_{\ell - 2}} E_{i_{\ell - 1} i_{\ell})^T} \partial^{(i_1} \dots \partial^{i_{\ell - 2}} E^{i_{\ell - 1} i_{\ell})^T} \delta^{(3)}\left(x - x_0\right) \\
&\hspace{0.5cm} + \sum_{\ell = 2}^{\infty} \frac{\lambda_{\ell}^{(C_B)}}{2\ell!} \int \dd{t} \dd[3]{x} \partial_{(i_1} \dots \partial_{i_{\ell - 2}} B_{i_{\ell - 1} i_{\ell})^T} \partial^{(i_1} \dots \partial^{i_{\ell - 2}} B^{i_{\ell - 1} i_{\ell})^T} \delta^{(3)}\left(x - x_0\right),
\end{aligned}
\end{equation}
where $E_{ij}$ and $B_{ij}$ are the electric and magnetic components of the Weyl tensor. We couple to the Weyl tensor only since couplings to the Ricci tensor and scalar can be removed through field redefinitions. $E_{ij}$ and $B_{ij}$ are introduced in the context of the Kaulza--Klein decomposition in \autoref{a:KKDecomp} with their expressions in terms of Kalulza--Klein variables given in \autoref{eq:EWeyl} and \autoref{eq:BWeyl}. As before, the point object is taken to be at rest and at the origin. Since we are working in the static limit, all time derivatives are set to zero. There are two Love numbers that encode the finite-size effect of tidal deformability, the gravito-electric Love number $\lambda^{(C_E)}$ and $\lambda^{(C_B)}$. It is worth highlighting that we are working in the limit where tidal perturbations behave as if we are strictly in flat space. There are, in principle, additional $r_s$ corrections but they are sub-leading and enter in higher orders in perturbation theory of the tidal field.  

With that said, the next step is to simplify the action. In the vector portion of the action, we once again recognize that $F_{ij}$ obeys the Bianchi identity. Hence, we can treat it as an independent field and add $\chi$ as a Lagrange multiplier, just as in \autoref{eq:EMAuxiliary}. Additionally, in the Regge--Wheeler gauge of \autoref{eq:RWGauge}, the action reduces to
\begin{equation}
\label{eq:Spin2Bulk}
\begin{aligned}
S_{\text{bulk}} &= -\frac{1}{2} \int \dd{t} \dd[3]{x} \left(\partial_i \delta \phi \partial^i \delta \phi - \frac{1}{2} F^{ij} F_{ij} + \chi \epsilon^{ijk} \partial_i F_{jk}\right) \\
&\hspace{0.4cm} + \frac{1}{4} \int \dd{t} \dd{r} \dd{\theta} \dd{\varphi} \sqrt{\gamma} \left(\frac{r^2}{2}\left(\partial_r \mathcal{K}\right)^2 + r \mathcal{K} \partial_r h_{rr} + 2 h_{rr}^2 + D_a h_{rr} D^a \mathcal{K}\right)  \\
&\hspace{0.4cm} + \frac{1}{2}\int \dd{t} \dd{r} \dd{\theta} \dd{\varphi} \sqrt{\gamma} \frac{1}{r^2}\left(\left(D^2 \mathcal{H}_O\right)^2 - 2 D_a \mathcal{H}_O D^a \mathcal{H}_O\right).
\end{aligned}
\end{equation}
Notice that in the flat space limit, $h_{rr}$ and $\mathcal{K}$ completely decouple from $\delta \phi$.\footnote{From Eq. \eqref{eq: spin 2 mixing}, we see that the mixing between the 3D metric perturbations and $\delta \phi$ becomes negligible at distances $r$ such that $|\partial_i \bar \phi| \ll 1/r$. This condition implies $r_s / r \ll 1$, and therefore the decoupling limit coincides with the large distance limit. This is somewhat reminiscent of what happens in single-field slow-roll inflation. There, metric perturbations decouple from the inflaton fluctuations $\delta \phi$ when at time scales $t$ such that $\partial_t \bar \phi \ll 1/t$, and this decoupling limit coincides with the de Sitter limit.} We can therefore safely integrate out these fields as well as $F_{ij}$ yielding 
\begin{equation}
\label{Sbulkspin2}
S_{\text{bulk}} = -\frac{1}{2}\int \dd{t} \dd[3]{x} \left(\partial_i \delta \phi \partial^i \delta \phi + \partial_i \chi \partial^i \chi\right). 
\end{equation}
On the other hand, in terms of these variables, the Love number action simplifies to 
\begin{equation}
\label{Slovespin2}
\begin{aligned}
S_{\text{Love}} &= \sum_{\ell = 2}^{\infty} \frac{\lambda_{\ell}^{(C_E)}}{2\ell!} \int \dd{t} \dd[3]{x} \partial_{(i_1} \dots \partial_{i_{\ell})^T} \delta\phi \partial^{(i_1} \dots \partial^{i_{\ell})^T} \delta\phi \delta^{(3)}\left(x - x_0\right) \\
&\hspace{0.5cm} + \sum_{\ell = 2}^{\infty} \frac{\lambda_{\ell}^{(C_B)}}{2\ell!} \int \dd{t} \dd[3]{x} \partial_{(i_1} \dots \partial_{i_{\ell})^T} \chi \partial^{(i_1} \dots \partial^{i_{\ell})^T} \chi \delta^{(3)}\left(x - x_0\right) \, ,
\end{aligned}
\end{equation}
where we have taken the large distance limit, dropping terms suppressed by powers of $r_s$.
The total action is once again the sum of two scalar bulk and worldline actions, with one term corresponding to the even-parity degree of freedom and one corresponding to the odd-parity degree of freedom.

\autoref{Sbulkspin2} and \autoref{Slovespin2} raise a natural question: what symmetries are respected by $S_{\rm bulk}$ but violated by $S_{\rm Love}$? 
A black hole would be an object for which imposing such symmetries are appropriate. Alternatively, such symmetries are broken for a generic object for which the Love numbers are non-zero.

Just as in the cases of spin-0 and spin-1, a reasonable guess for the correct EFT symmetries is to take the long distance limit of the known geometric symmetries for a black hole (\autoref{Symm_spin2}; see paragraph below \autoref{R0Rscalar}). For instance,
the corresponding symmetry generated by $K_3$ is (in the large $r$ limit):
\begin{equation}
    \delta_{\xi} \delta \phi = r^2\cos\theta \, \partial_r \delta \phi + r\partial_\theta\left(\sin\theta \delta \phi\right),
\end{equation}
and similarly so for $\chi$. 
The full set of special conformal symmetries generated by the melodic CKVs $K_i$ are, in the long distance limit, expressible in Cartesian coordinates by replacing $\phi$ in \autoref{eq:scalarspecialconformal2} by $\delta \phi$ or $\chi$. It can be shown these special conformal transformations are symmetries of $S_{\rm bulk}$, but not of $S_{\rm Love}$ (see \autoref{eftspin0} for a discussion). 

\section{Discussion}
\label{discuss}
In this paper, we study the geometric origin for the vanishing of static spin 1 and spin 2 Love numbers in a Schwarzschild background. We demonstrate that for both the spin 1 and 2 static perturbations, the even and odd sectors respectively have their own SO(3,1) symmetries. In addition, an SO(2) symmetry connects the two sectors, essentially an electric-magnetic duality. This replicates the story known for spin 0 static perturbations. Representation theory of SO(3,1) tells us there is a ladder structure relating perturbations at different values of the angular momentum $\ell$. Vanishing of the black hole Love numbers follows from the nature of the ground state solution ($\ell = 1$ for spin 1, $\ell = 2$ for spin 2), and the higher $\ell$ solutions built from it by climbing the ladder: that all solutions that are regular at the horizon have no tidal tail far away from the black hole. An advantage of our approach is that we were able to find new ladder structures that act directly on the gauge field variables, which complements the already known ladders for the Newman-Penrose variables \cite{Hui2022}. We can also readily write down the corresponding conserved Noether current, because of our formulation of the symmetries at the action level. Moreover, it was important that we were able to reduce the action describing the physical degrees of freedom to two scalars living in 3D effective metrics, which allows us to leverage our knowledge of melodic CKV symmetries for scalars. As a byproduct of this analysis, we found new methods for computing all melodic CKVs, such as noting that the space of melodic CKVs in two metrics related by a conformal factor are the same so long as that conformal factor is proportional to the ratio of the Ricci scalars of the respective metrics.

The large $r$ limit of (the ``boost" part of) the geometric symmetries are the standard special conformal transformations in 3D flat space. It can be shown they are symmetries of the bulk action, but not of the Love number terms on the worldline, in the static limit.
It's worth emphasizing that the special conformal symmetries apply strictly in the static limit. Thus, we do not expect them to protect black holes from quantum corrections generating non-zero Love numbers, i.e. even if the external tidal field is static, quantum loops can involve non-static field configurations. 

Several issues are worth further investigation. 
First, as discussed in \cite{Kehagias:2024rtz,Combaluzier-Szteinsznaider:2024sgb,Gounis:2024hcm}, some of the symmetries of the linear theory can persist into the non-linear theory, provided the right variables are used. It would be interesting to extend their work, which assumed axisymmetry, to more general configurations. 
The choice of variables we found which makes the linear symmetries manifest might give useful hints for this exploration. 

Second, the static Love numbers also vanish for a Kerr black hole.
For spin 0 perturbations, it is possible to derive a 3D effective metric with two melodic CKVs directly from the general formula \autoref{eq:GeneralEffMetric}. 
It would be interesting to work out the same for spin 1 and 2 perturbations, for which care must be taken to identify the correct even and odd sectors. Recently, hidden symmetries were uncovered for axisymmetric and static perturbations in the Teukolsky equation \cite{Lupsasca2025}.

Third, while we focus on black holes, neutron star Love numbers are also very interesting. They are known to be non-zero, and in fact there appears to be certain universal relations governing different properties of the neutron star, including its Love numbers, quadrupole moment and moment of inertia \cite{Yagi:2013}. Is there some sense in which the SO(3,1) symmetries are weakly broken by a neutron star? If so, can it shed light on some of these universal relations? 

Lastly, the vanishing of static Love numbers is particular to four-dimensional black holes. In higher dimensions, the Love numbers only vanish when $\hat{\ell} = \ell/\left(d - 3\right)$ takes an integer value \cite{Hui2021}.
(Here, $d$ is the number of spacetime dimensions.) As is shown in \cite{Berens:2025}, there exist ladder structures for higher dimensional black holes, including the Tangherlini and Myers--Perry solutions. However, a geometric understanding for the origin of those ladders is still lacking. It is possible that such a geometric origin is based on symmetries generated by CKVs as was shown here in four spacetime dimensions.

\paragraph{Acknowledgments.} We would like to thank Austin Joyce, Massimiliano Riva, Luca Santoni, Alessandro Podo, and Adam Solomon for many useful discussions throughout the course of this project. The work of RP is supported by the US Department of Energy grant DE-SC0010118. The work of LH, DM and JS is supported by the US Department of Energy grant DE-SC011941. The work of RB is supported by the NSF CAREER award PHY-2340457 and the Simons Foundation award SFI-MPS-BH-00012593.

\newpage

\appendix

\section{Melodic Conformal Killing Vectors}
\label{MCKV}

\subsection{The Melodic Condition}

\label{a:MCKVs1} The melodic condition arises from the question of whether there are conformal symmetries of a scalar action 
\begin{equation}
\label{eq:ScalarActionCoupling}
S = - \frac{1}{2}\int\dd[d]{x} \sqrt{g} \left(g^{ij}\nabla_{i}\phi\nabla_{j}\phi+\alpha R\phi^{2}\right)
\end{equation}
with arbitrary coupling $\alpha$ to the Ricci scalar (i.e. whose coupling is \textit{not} necessarily the conformal coupling, $\alpha_{c}=(d-2)/[4(d-1)]$ for $d$ spacetime dimensions). This is particularly pertinent to this work since we are typically interested in the minimally coupled scalar action $\alpha = 0$ and, in one case, we have $\alpha = 2$.

We consider the variation of \autoref{eq:ScalarActionCoupling} under a conformal transformation generated by $\xi^i$ such that the length element transforms as
\begin{equation}
    \dd{x'}^2 = \Omega^2\dd{x}^2,
\end{equation}
with $\Omega\left(x\right)$ the conformal factor. To first order, $\dd{x'}^2 = g_{ij}\dd{x}^i \dd{x}^j + 2\left(\ln \Omega\right) g_{ij} \dd{x}^i \dd{x}^j$. This results in:
\begin{equation}
    \nabla_i \xi_j + \nabla_j \xi_i = -2\ln\Omega \: g_{ij}.
\end{equation}
Taking the trace of both sides tells us that $\nabla_i \xi^i = -d\ln\Omega$ leading to the conformal Killing equation
\begin{equation}
\label{eq:ConformalKillingEquation}
    \nabla_i \xi_j + \nabla_j \xi_i = \frac{2}{d}\left(\nabla_k \xi^k\right)g_{ij}.
\end{equation}

Under a conformal transformation a scalar field $\phi\left(x\right)$ transforms as 
\begin{equation}
    \phi'\left(x'\right) = \Omega^{-\Delta}\phi\left(x\right),
\end{equation}
with conformal weight $\Delta=\left(d-2\right)/2$. Infinitesimally, this is
\begin{equation}
\label{eq:VariationOfPhi}
    \delta_{\xi}\phi = \xi^{i}\nabla_{i}\phi + \frac{d - 2}{2d}\left(\nabla_{i}\xi^{i}\right)\phi.
\end{equation}
Substituting this into the action and rearranging yields
\begin{equation}
\begin{aligned}\delta_{\xi}S =-\int\dd[d]{x}\sqrt{g}\bigg(&\nabla_{i}\left(\frac{1}{2}\xi^{i}\nabla_{j}\phi\nabla^{j}\phi+\frac{n-2}{4n}\phi^{2}\nabla^{i}\nabla_{j}\xi^{j}\right)-\frac{n-2}{4n}\Box\left(\nabla\cdot\xi\right)\phi^{2}\\
 &+\frac{1}{2}\alpha\nabla_{i}\left(R\xi^{i}\phi^{2}\right)-\frac{\alpha}{n}\left(R\left(\nabla_{j}\xi^{j}\right)+\frac{n}{2}\xi^{i}\nabla_{i}R\right)\phi^{2}\bigg).
\end{aligned}
\end{equation}
In order to demonstrate a symmetry of the action we would need to demonstrate a relationship between the last term on the first line and last term on the last line. Indeed, this relation can be shown (see \cite{Berens2023}) to be
\begin{equation}
\Box\nabla_{i}\xi^{i} = -\frac{1}{d-1}\left(R\nabla_{i}\xi^{i} + \frac{d}{2}\xi^{i}\nabla_{i}R\right).
\end{equation}
Using this result and integrating out the total derivatives leaves
us with
\begin{equation}
\label{eq:ActionMelodic}
\delta_{\xi}S = -\frac{d-1}{d}\left(\alpha-\frac{d-2}{4\left(d-1\right)}\right)\int \dd[d]{x} \sqrt{g} \left(\Box\left(\nabla\cdot\xi\right)\right)\phi^{2}.
\end{equation}
As one would expect, this will be zero if $\alpha$ is equal to the value of the conformal coupling, $\alpha_{c}=(d-2)/[4(d-1)]$; however, we also find an additional symmetry for arbitrary coupling for a subset of conformal Killing vectors obeying the condition
\begin{equation}
\label{eq:MelodicCondition}
\Box\nabla_{j}\xi^{j}=0.
\end{equation}
Conformal Killing vectors that obey this condition are known as \textit{melodic} \cite{Berens2023}.

That such a symmetry exists for a scalar field with arbitrary coupling to the Ricci scalar raises the question as to whether there is a higher spin analogue of the melodic condition, in particular for spin 1 and spin 2. The answer to the question appears to be no. It is easy to see how problems arise when higher spins are considered; the infinitesimal variation of a field $\phi_{I}$ of arbitrary spin in a flat background is given by\footnote{See Hugh Osborn's notes on Conformal field theory \href{https://www.damtp.cam.ac.uk/user/ho/CFTNotes.pdf}{here}.}
\begin{equation}
\delta_{\xi}\phi_{I} = \xi^i \partial_i \phi_I + \frac{d - 2}{2d}\left(\partial_i \xi^i\right)\phi_I - \frac{1}{4}\left(\partial_i \xi_j - \partial_j \xi_i\right)\left(\mathcal{S}^{ij}\right)_{I}^{\ \ J}\phi_{J},
\end{equation}
where $\mathcal{S}^{ij}$ are the appropriate spin matrices for the specific field under consideration satisfying
\begin{equation}
\comm{\mathcal{S}_{ij}}{\mathcal{S}_{kl}} = \delta_{ik}\mathcal{S}_{jl} - \delta_{jk} \mathcal{S}_{il} - \delta_{il}\mathcal{S}_{jk} + \delta_{jl}\mathcal{S}_{ij}.
\end{equation}
We thus see that for higher spins, there are contractions involving the field itself. A melodic condition for higher spins would therefore need to be a condition on the field as well, rather than just a property of the background.

\subsection{Counting and Finding Melodic Conformal Killing Vectors}

\label{a:MCKVs2} A feature of melodic conformal Killing vectors is that, upon rescaling the metric by its Ricci scalar, they become true Killing vectors of the rescaled space. That is, if $\xi^{i}$ is a melodic conformal Killing vector in a spacetime with metric $g_{ij}$, then it is a Killing vector of the metric 
\begin{equation}
\tilde{g}_{ij}=RL_{0}^{2}g_{ij},
\end{equation}
where $L_{0}$ is an arbitrary length scale and $R$ is the Ricci scalar of $g_{ij}$. The number of melodic conformal Killing vectors is then equal to the number of Killing vectors in a spacetime whose metric is scaled by the Ricci scalar.

That this is the case follows from the identity \cite{Berens2023}
\begin{equation}
\Box\nabla_{i}\xi^{i}=-\frac{1}{d-1}\left(R\nabla_{i}\xi^{i}+\frac{d}{2}\xi^{i}\nabla_{i}R\right),\label{eq:F3id}
\end{equation}
which is true for any conformal Killing vector. Recalling the melodic condition \autoref{eq:MelodicCondition}, for a melodic conformal Killing vector this identity becomes
\begin{equation}
\label{eq:MelodicConditionRicci}
R\nabla_{i}\xi^{i} + \frac{d}{2}\xi^{i}\nabla_{i}R = 0.
\end{equation}
We now assume that we are in a transformed space and that $\xi^{i}$ is a Killing vector for this space. Its divergence is therefore $0$. If we were to bring it back to the unscaled space, then we would need to use the connection $C_{\hphantom{i}ij}^{i}$ given by \cite[(D.3)]{Wald1984},
\begin{equation}
\begin{aligned}\widetilde{\nabla}_{i}\xi^{i} & =\nabla_{i}\xi^{i}+C_{\hphantom{i}ij}^{i}\xi^{j}\\
 & =\nabla_{i}\xi^{i}+\left(\delta_{\hphantom{i}i}^{i}\nabla_{j}\ln\Omega+\delta_{\hphantom{i}j}^{i}\nabla_{i}\ln\Omega - g_{ij}\nabla^{i}\ln\Omega\right)\xi^{\nu}\\
 & =\nabla_{i}\xi^{i} + d\xi^{i}\nabla_{i}\ln\Omega.
\end{aligned}
\label{eq:ConfDiv}
\end{equation}
Setting this to $0$ since $\xi^{i}$ is a Killing vector in the transformed space gives
\begin{equation}
\Omega^{2}\nabla_{i}\xi^{i}+\frac{d}{2}\xi^{i}\nabla_{i}\Omega^{2}=0.
\end{equation}
Comparing with \autoref{eq:MelodicConditionRicci}, we see that if $\xi^{i}$ is a melodic Killing vector, then the above equation is true if $\Omega^{2}=RL_{0}^{2}$ for a constant $L_{0}$ of one length dimension. Similarly, if the transformation factor $\Omega^2$ is $RL_{0}^{2}$, we automatically find that $\xi^{i}$ is a conformal Killing vector. As a result, if $\xi^{i}$ is a Killing vector of a space $\tilde{g}_{ij}$, it is a melodic Killing vector of $g_{ij}$ and if $\xi^{i}$ is a melodic Killing vector of $g_{ij}$, it is a Killing vector of $\tilde{g}_{ij}$. In other words, the melodic CKVs of a metric $g_{ij}$ are the KVs of $\tilde{g}_{ij}=RL_{0}^{2}g_{ij}$.

This is a powerful result for our purposes since we will mostly focus on three-dimensional spaces for which there exists a detailed algorithm for deriving the number of Killing vectors. In particular, the authors of \cite{Kruglikov2018} were able to derive an algorithm that counts the exact number of Killing vectors for any 3D Riemannian manifold.

\subsection{Scaling Transformations that Preserve Melodicity}
\label{a:MCKVs3} 

Given the prevalence of effective metrics conformally
related to the scalar effective metric in this work, it is useful to ask is whether or not scaling transformations preserve melodicity. That is whether, upon conformal transformation, melodic CKVs of the original metric are also melodic CKVs of the transformed metric. If melodicity is preserved, then we immediately know the melodic CKVs of the transformed space. We begin by scaling the metric by $\Omega^{2}$ once again, meaning $\tilde{g}_{ij} = \Omega^2 g_{ij}$, and writing down the melodic condition in the scaled space,
\begin{equation}
\widetilde{\vphantom{\rule{0pt}{1.4ex}}\Box} \tilde{\nabla}_{i}\xi^{i}=-\frac{1}{d-1}\left(\tilde{R}\tilde{\nabla}_{i}\xi^{i}+\frac{d}{2}\xi^{i}\tilde{\nabla}_{i}\tilde{R}\right).
\end{equation}
Transforming back to the original space, we find
\begin{equation}
\widetilde{\vphantom{\rule{0pt}{1.4ex}}\Box} \tilde{\nabla}_{i}\xi^{i}=\Omega^{-2}\left(\Box\nabla_{i}\xi^{i}+2\left(\nabla_{i}\xi^{i}+\frac{d}{2}\xi^{i}\nabla_{i}\right)\left(\Box\ln\Omega+\frac{d-2}{2}\nabla_{j}\ln\Omega\nabla^{j}\ln\Omega\right)\right).
\end{equation}
This directly implies that melodicity is preserved when 
\begin{equation}
\Box\ln\Omega+\frac{d-2}{2}\nabla_{j}\ln\Omega\nabla^{j}\ln\Omega=\alpha R
\end{equation}
for some constant $\alpha$, since that means the above equation can be rewritten using \autoref{eq:F3id} as 
\begin{equation}
\widetilde{\vphantom{\rule{0pt}{1.4ex}}\Box} \tilde{\nabla}_{i}\xi^{i}=\Omega^{-2}\left(1-2\alpha\left(d-1\right)\right)\Box\nabla_{i}\xi^{i}.
\end{equation}
Notice that this also means, due to \cite[eq. (D.9)]{Wald1984}, that 
\begin{equation}
\tilde{R}=\Omega^{-2}\left(1-2\alpha\left(d-1\right)\right)R.
\end{equation}
Thus, the value for $\alpha$ which ensures all CKVs become melodic is one in that maps you to a spacetime with vanishing Ricci scalar. Moreover, melodicity is preserved when 
\begin{equation}
\tilde{R}\propto\Omega^{-2}R.
\end{equation}

\subsection{The Noether Current}
\label{a:MCKVs4} 

Since melodic conformal transformations constitute symmetries of a scalar action with arbitrary coupling to the Ricci scalar, it is natural to ask what the associated Noether current is. Noether's theorem tells us that for a symmetry variation of the field $\delta_{s}\phi$, there is a conserved current
\begin{equation}
\label{eq:GeneralNoetherCurrent}
J^{i}=\pdv{\mathcal{L}}{\left(\nabla_{i}\phi\right)}\delta_{s}\phi-\Lambda^{i},
\end{equation}
where we take the Lagrangian $\mathcal{L}$ to not include the integration measure $\sqrt{g}$ and $\Lambda^{i}$ is defined such that $\delta_{s}\mathcal{L}=\nabla_{i}\Lambda^{i}$. $\Lambda^i$ is determined from the total derivative that was integrated out in \autoref{eq:ActionMelodic}, 
\begin{equation}
\Lambda^{i} = -\frac{1}{2}\left(\xi^{i}\nabla_{j}\phi\nabla^{j}\phi+\frac{d-2}{2d}\phi^{2}\nabla_{i}\nabla_{j}\xi^{j}+\alpha R\xi^{i}\phi^{2}\right),
\end{equation}
where $\alpha$ is the coupling of the scalar field to the Ricci scalar in the action. Combining this with the variation of the scalar field given in equation \autoref{eq:VariationOfPhi}, we obtain the Noether current 
\begin{equation}
J^{i} = -\xi^{j}\nabla_{j}\phi\nabla^{i}\phi - \frac{d-2}{4d}\nabla_{j}\xi^{j}\nabla^{i}\phi^{2} + \frac{1}{2}\xi^{i}\nabla_{j}\phi\nabla^{j}\phi + \frac{d-2}{4d}\phi^{2}\nabla^{i}\nabla_{j}\xi^{j} + \frac{\alpha}{2}R\xi^{i}\phi^{2}.
\end{equation}

We now want to demonstrate that this current is actually conserved.
Applying $\nabla_{i}$ on both sides, we obtain 
\begin{equation}
\begin{aligned}\nabla_{i}J^{i} & =-\nabla_{i}\xi^{j}\nabla_{j}\phi\nabla^{i}\phi-\xi^{j}\nabla_{i}\nabla_{j}\phi\nabla^{i}\phi-\xi^{j}\nabla_{j}\phi\Box\phi\\
 & \hspace{0.5cm}-\frac{d-2}{2d}\nabla_{i}\nabla_{j}\xi^{j}\left(\phi\nabla^{i}\phi\right)-\frac{d-2}{2d}\nabla_{j}\xi^{j}\nabla_{i}\phi\nabla^{i}\phi-\frac{d-2}{2d}\nabla_{j}\xi^{j}\left(\phi\Box\phi\right)\\
 & \hspace{0.5cm}+\frac{1}{2}\nabla_{i}\xi^{i}\nabla_{j}\phi\nabla^{j}\phi+\frac{1}{2}\xi^{i}\nabla_{i}\nabla_{j}\phi\nabla^{j}\phi+\frac{1}{2}\xi^{j}\nabla_{j}\phi\nabla_{i}\nabla^{j}\phi\\
 & \hspace{0.5cm}+\frac{d-2}{4d}\left(\nabla_{i}\phi^{2}\right)\nabla_{i}\nabla_{j}\xi^{j}+\frac{\alpha}{2}\left(\xi^{i}\nabla_{i}R\right)\phi^{2}+\frac{\alpha}{2}R\phi^{2}\nabla_{i}\xi^{i}+\frac{\alpha}{2}R\xi^{i}\nabla_{i}\phi^{2},
\end{aligned}
\end{equation}
where we have imposed the melodic condition, $\Box\nabla_{j}\xi^{j} = 0$. In simplifying this, it is useful to note that the conformal Killing equation gives
\begin{equation}
-\nabla^{(i}\xi^{j)}\nabla_{j}\phi\nabla_{i}\phi=-\frac{1}{2}\left(\nabla^{i}\xi^{j}+\nabla^{j}\xi^{i}\right)\nabla_{i}\phi\nabla_{j}\phi=-\frac{1}{d}\nabla_{i}\xi^{i}\nabla_{j}\phi\nabla^{j}\phi.
\end{equation}
Using this, as well as relabeling some indices, we find
\begin{equation}
\nabla_{i}J^{i}=-\xi^{j}\nabla_{j}\phi\Box\phi-\frac{d-2}{2d}\phi\nabla_{j}\xi^{j}\Box\phi+\frac{\alpha}{2}\xi^{i}\phi^{2}\nabla_{i}R+\frac{\alpha}{2}R\phi^{2}\nabla_{i}\xi^{i}+\frac{\alpha}{2}R\xi^{i}\nabla_{i}\phi^{2}.\label{eq:covJ}
\end{equation}
If we insert the equation of motion $\Box\phi = -\alpha R\phi$,
this reduces to 
\begin{equation}
\nabla_{i}J^{i}=\frac{\alpha}{d}\phi^{2}\left(R\nabla_{i}\xi^{i}+\frac{d}{2}\xi^{i}\nabla_{i}R\right)
\end{equation}
However, if the melodic condition is satisfied then the term in the parentheses is also $0$ by \autoref{eq:F3id}.

\section{Transformation of the Love Number Operators}
\label{breakingSCT}

In this Appendix, we show that while $\delta {\cal L}_{\rm Love}$
is not a total derivative for special conformal transformation, it
is a total derivative for rotation.
Specifically, for rotation:
\begin{equation}
  \delta_\xi \phi = \xi^i \partial_i \phi \quad , \quad
  \xi^i = \epsilon^{ijk} c_j x_k \, ,
\end{equation}
where $\vec c$ is a constant vector specifying the axis of rotation.
We will work out explicitly the case of $\ell=0$.
The other cases can be worked out in a similar way.

For $\ell=0$:
\begin{subequations}
    \begin{align}
        \delta \mathcal{L}_{\text{Love}} &= \lambda_0 \xi^i \partial_i \phi \: \phi \delta^{(3)}\left(\vec{x}\right),\\
        &= \partial_i\left(\frac{\lambda_0}{2} \epsilon^{ijk} c_j x_k \phi^2 \delta^{(3)}\left(\vec{x}\right)\right) - \frac{\lambda_0}{2} \partial_i\left(\epsilon^{ijk} c_j x_k\right) \phi^2 \delta^{(3)}\left(\vec{x}\right) - \frac{\lambda_0}{2} \epsilon^{ijk} c_j x_k \phi^2 \partial_i \delta^{(3)}\left(\vec{x}\right).
    \end{align}
\end{subequations}
The second term on the right obviously vanishes.
The third term on the right also vanishes if one
writes $\delta^{(3)} (\vec x) = \int e^{i \vec q \cdot \vec x} d^3 q /
(2\pi)^3$. This is because the combination
$\epsilon^{ijk} x_k q_i = (\vec x \times \vec q)^j$ can be rewritten
as $(\vec x \times
\vec q_\perp)^j$, if one decomposes $\vec q$ into components parallel
and perpendicular to 
$\vec x$: $\vec q \equiv \vec q_{\parallel} + \vec q_\perp$.
The 2D integral over $\vec q_\perp$ yields zero because
the integrand is odd in $\vec q_\perp$ (keeping in mind $e^{i \vec q
  \cdot \vec x}$ is independent of $\vec q_\perp$).
The case of $\vec x = 0$ has to be treated separately, but
again the oddness under $\vec q \rightarrow -\vec q$ guarantees
the momentum integral yields zero.
Therefore we conclude $\delta {\cal L}_{\rm Love}$ takes the
form of a total derivative under rotation:
$\delta {\cal L}_{\rm Love} = \partial_i \Lambda^i_{\rm Love}$.
Standard arguments tell us the Noether current is
$J^i \equiv \xi^j\partial_j \phi
\left( \partial [{\cal L}_{\rm bulk} + {\cal L}_{\rm Love}] / \partial
  (\partial_i \phi) \right) - \Lambda^i - \Lambda^i_{\rm Love}$,
where $\Lambda^i$ is given by \autoref{deltabulk} for rotational
$\xi^i$. It can be checked that its conservation is consistent with
the scalar equation of motion \autoref{scalareomEFT}.
As remarked in \Cref{s:StaticScalar}, one must take care not to
treat $\Lambda^i_{\rm Love}$ as vanishing; otherwise, the consistency
check would not have worked out.

\section{The Kaluza--Klein Decomposition}
\label{a:KKDecomp}
In this section we seek to review the Kaluza--Klein decomposition, which is often associated with dimensional reductions in cases where the metric does not depend on one of the coordinates. Note that we will use $\alpha, \beta, \gamma$ $\delta$, etc. for 4D tetrad indices, while we use $\mu, \nu, \rho$, $\sigma$, etc. for 4D coordinate basis indices. Meanwhile, $a$, $b$, $c$, and $d$ will be used for 3D tetrad indices, while $i$, $j$, $k$, etc. will be used for 3D coordinate basis indices.

The decomposition begins with a convenient parameterization of the metric, aptly called the \textit{Kaluza--Klein metric},
\begin{equation}
\label{eq:KKMetric}
    \dd{s}^2 = -e^{-\phi}\left(\dd{t} + A_i \dd{x}^i\right)^2 + e^{\phi}g_{3\: ij} \dd{x}^i \dd{x}^j.
\end{equation}
An introduction to this is presented in the \href{http://people.tamu.edu/~c-pope/ihplec.pdf}{lecture notes} by C.N. Pope. Our goal will be to write down the Riemann tensor in terms of these Kaluza--Klein variables. For computational simplicity, define
\begin{equation}
    \tilde{g}_{ij} = e^{\phi}g_{3\: ij}.
\end{equation}

We define the tetrad $\hat{e}^\alpha$ as follows
\begin{equation}
    \hat{e}^0 = e^{-\phi/2}\left(\dd{t} + A_i\dd{x}^i\right) \qquad \text{and} \qquad \hat{e}^a = \tilde{e}^a_{\hphantom{a}i}\dd{x}^i,
\end{equation}
where $\tilde{e}^a_{\hphantom{a}i}$ is the tetrad associated to $\tilde{g}_{ij}$. The tetrad formulation is particularly useful since the presence of the generic 3D tetrad will allow us to more easily express 4D objects, such as the Riemann tensor, in terms of equivalent 3D objects. Note that all spatial indices will be raised and lowered with $\tilde{g}_{ij}$, meaning
\begin{equation}
    A^i = \tilde{g}^{ij}A_j \qquad \text{and} \qquad A^a = \tilde{e}^a_{\hphantom{a}i} A^i.
\end{equation}
Now we can compute the spin-connection,
\begin{equation}
    \omega^\alpha_{\hphantom{\alpha}\beta} = \omega^{\hphantom{\mu}\alpha}_{\mu\hphantom{\alpha}\beta}\dd{x}^\mu = \omega^{\hphantom{\gamma}\alpha}_{\gamma\hphantom{\alpha}\beta}\hat{e}^\gamma.
\end{equation}
The spin-connection is used to define the covariant derivative in tetrad indices for a mixed tensor $X^a_{\hphantom{a}b}$,
\begin{equation}
\label{eq:covdtet}
    \widetilde{\nabla}_c X^a_{\hphantom{a}b} = \partial_c X^a_{\hphantom{a}b} + \tilde{\omega}^{\hphantom{c}a}_{c\hphantom{a}d} X^d_{\hphantom{d}b} - \tilde{\omega}^{\hphantom{c}d}_{c\hphantom{d}b}X^a_{\hphantom{a}d},
\end{equation}
where $\tilde{\omega}^a_{\hphantom{a}b}$ is the spin connection associated with $\tilde{e}^a$. We can compute these spin-connections through the torsion-free condition,
\begin{equation}
    T^\alpha = \dd{\hat{e}}^\alpha + \omega^\alpha_{\hphantom{\alpha}\beta}\wedge \hat{e}^\beta.
\end{equation}
The choice of spin-connection that ensures this is zero is 
\begin{equation}
    \omega^0_{\hphantom{0}a} = -\frac{1}{2}\partial_a \phi \: \hat{e}^0 + \frac{1}{2} e^{-\phi/2} F_{ab} \hat{e}^b,
\end{equation}
where $F_{ab} = \partial_a A_b - \partial_b A_a$, and 
\begin{equation}
    \omega^a_{\hphantom{a}b} = \tilde{\omega}^a_{\hphantom{a}b} + \frac{1}{2} e^{-\phi/2} F^a_{\hphantom{a}b}\hat{e}^0.
\end{equation}
From here, we can compute the curvature two-form,
\begin{equation}
    \mathcal{R}^\alpha_{\hphantom{\alpha}\beta} = \dd{\omega}^\alpha_{\hphantom{\alpha}\beta} + \omega^\alpha_{\hphantom{\alpha}\gamma} \wedge \omega^\gamma_{\hphantom{\gamma}\beta}. 
\end{equation}
Recall that we can relate the components of the curvature two-form to the components of the Riemann tensor
\begin{equation}
\mathcal{R}^\alpha_{\hphantom{\alpha}\beta} = \frac{1}{2}R^\alpha_{\hphantom{\alpha}\beta\gamma\delta} \hat{e}^\gamma \wedge \hat{e}^\delta.
\end{equation}
The Riemann tensor components are
\begin{equation}
    \begin{aligned}
        R^0_{\hphantom{0}a0b} &= \frac{1}{2}\widetilde{\nabla}_b \widetilde{\nabla}_a \phi - \frac{1}{4}\widetilde{\nabla}_a \phi \widetilde{\nabla}_b \phi + \frac{1}{4}e^{-\phi} F_{ac} F^c_{\hphantom{c}b}.\\
        R^0_{\hphantom{0}abc} &= \frac{1}{2}e^{-\phi/2}\left(\widetilde{\nabla}_a F_{bc} - F_{bc}\widetilde{\nabla}_a \phi + \frac{1}{2}F_{ab} \widetilde{\nabla}_c \phi - \frac{1}{2}F_{ac}\widetilde{\nabla}_b \phi\right),\\
        R^a_{\hphantom{a}bcd} &= \tilde{R}^a_{\hphantom{a}bcd} + \frac{1}{2}e^{-\phi}\left(F^a_{\hphantom{a}b} F_{cd} + \frac{1}{2} F^a_{\hphantom{a}c} F_{bd} - \frac{1}{2}F^a_{\hphantom{a}d}F_{bc}\right)
    \end{aligned}
\end{equation}
Note that to simplify this expression further, we made use of the Bianchi identity, $\widetilde{\nabla}_{[a}F_{bc]} = 0$. 

In order to find the Riemann tensor for our original metric, \autoref{eq:KKMetric}, we'll need to insert $\tilde{g}_{ij} = e^{\phi} g_{3\:ij}$. The situation is slightly more complicated than in the coordinate basis since, in the tetrad basis, spacetime dependence is contained not within the metric, but rather within the tetrads themselves. In other words,
\begin{equation}
    e^{\phi} g_{ij} = \eta_{ab} \tilde{e}^a_{\hphantom{a}i} \tilde{e}^b_{\hphantom{b}j} = e^{\phi} \eta_{ab} e^a_{\hphantom{a}i} e^b_{\hphantom{b}j} \rightarrow e^a_{\hphantom{a}i} = e^{-\phi/2} \tilde{e}^a_{\hphantom{a}i},
\end{equation}
where $e^a$ is the tetrad associated to $g_{3\:ij}$. Now let's study how vectors and one-forms transform. It is important to distinguish between these since, for example, raising the index on a one-form will include a metric implicitly. For a vector, $V$, and one-form $X$,
\begin{equation}
\begin{aligned}
    V^a &= \tilde{e}^a_{\hphantom{a}i} V^i \rightarrow e^{\phi/2} V^a \qquad & \qquad V_a &= \tilde{e}_a^{\hphantom{a}i} \left(g_{ij} V^j\right) \rightarrow e^{\phi/2} V_a,\\
    X_a &= \tilde{e}_a^{\hphantom{a}i} X_i \rightarrow e^{-\phi/2} X_a \qquad & \qquad X^a &= \tilde{e}^a_{\hphantom{a}i} \left(g^{ij} X_j\right) \rightarrow e^{-\phi/2} X^a.
\end{aligned}
\end{equation}
Contractions then transform as follows:
\begin{equation}
    V_a V^a \rightarrow e^{\phi} V_a V^a \qquad \text{and} \qquad X_a X^a \rightarrow e^{-\phi} X_a X^a.
\end{equation}
One other object that we will need to transform is the covariant derivative in the tetrad basis. We'll do that by solving for the spin-connection, through the zero torsion condition,
\begin{equation}
    \dd{e}^a = -\omega^a_{\hphantom{a}b} \wedge e^b.
\end{equation}
Then, using $e^a = e^{-\phi/2} \tilde{e}^a$,
\begin{equation}
    \dd{\left(e^{-\phi/2}\tilde{e}^a\right)} = -\frac{1}{2} \partial_b \phi \: e^b \wedge e^a - \tilde{\omega}^a_{\hphantom{a}b} \wedge e^b.
\end{equation}
Combining both of the previous equations,
\begin{equation}
    \tilde{\omega}^a_{\hphantom{a}b} \wedge e^b = \omega^a_{\hphantom{a}b} \wedge e^b + \frac{1}{2} \partial_b \phi\: e^a \wedge e^b.
\end{equation}
The spin connection that solves this and is antisymmetric in $a$ and $b$ is
\begin{equation}
\label{eq:transspincon}
    \tilde{\omega}^a_{\hphantom{a}b} = \omega^a_{\hphantom{a}b} + \frac{1}{2} \left(\delta^a_{\hphantom{a}c}\partial_b\phi - \eta_{bc}\partial^a \phi\right)e^c.
\end{equation}
It is important to note that $\tilde{\omega}^a_{\hphantom{a}b} = \tilde{\omega}^{\hphantom{c}a}_{c\hphantom{a}b} \tilde{e}^c$. We need to convert that to $e^c$, which amounts to multiplying both sides by $e^{-\phi/2}$, so
\begin{equation}
    \tilde{\omega}^{\hphantom{c}a}_{c\hphantom{a}b} = e^{-\phi/2} \omega^{\hphantom{c}a}_{c\hphantom{a}b} + \frac{1}{2} e^{-\phi/2}\left(\delta^a_{\hphantom{a}c} \partial_b\phi - \eta_{bc} \partial^a \phi\right).
\end{equation}
This overall factor has $e^{-\phi/2}$, which is in line with our previous rule that in a tensor defined with $n$ upper and $m$ lower indices, the transformation in the tetrad basis would pick up a factor of $e^{(n - m)\phi/2}$. The spin-connection, however, picks up an additional connection,
\begin{equation}
    C^{\hphantom{c}a}_{c\hphantom{a}b} = \frac{1}{2}\left(\delta_c^{\hphantom{c}a}\nabla_b \phi - \eta_{cb} \nabla^a \phi\right).
\end{equation}

We summarize what happens when we perform conformal transformations, $e \rightarrow \tilde{e} = \Omega e$, in the tetrad basis:
\begin{enumerate}[label=\textbf{\arabic*.}]
\item Vectors transform with a factor of $\Omega$ regardless of whether it is an upper or lower index, while one forms transform with a factor of $\Omega^{-1}$ regardless of whether it is an upper or lower index.
\item Mixed tensors transform in a consistent way. An $(n, m)$ tensor transforms with $\Omega^{(n - m)}$ regardless of the position of the indices.
\item Covariant derivatives transform according to the connection
\begin{equation}
    C^{\hphantom{c}a}_{c\hphantom{a}b} = \delta_c^{\hphantom{c}a}\nabla_b \ln\Omega - \eta_{cb} \nabla^a \ln\Omega.
\end{equation}
They will then include an overall factor based on the first two rules.
\end{enumerate}

With this set, it is straight-forward to transform the Riemann tensor components, the results of which are listed below:
\begin{equation}
    \begin{aligned}
        R^0_{\hphantom{0}a0b} &= \frac{1}{2}e^{-\phi}\nabla_b \nabla_a \phi - \frac{3}{4}e^{-\phi}\nabla_a \phi \nabla_b \phi + \frac{1}{4}e^{-\phi}\eta_{ab}\nabla_c \phi \nabla^c \phi + \frac{1}{4} e^{-3\phi} F_{ac} F^c_{\hphantom{c}b},\\
        R^0_{\hphantom{0}abc} &= \frac{1}{2}e^{-2\phi}\Big(\nabla_a F_{bc} - F_{bc}\nabla_a + \frac{1}{2} F_{ab}\nabla_c \phi - \frac{1}{2} F_{ac}\nabla_b \phi \\
        &\hspace{1.5cm} - F_{bc}\nabla_a \phi - \frac{1}{2} F_{ab}\nabla_b \phi + \frac{1}{2}F_{ab}\nabla_c \phi - \frac{1}{2}\eta_{ab} F_{cd}\nabla^d \phi + \frac{1}{2}\eta_{ac} F_{bd}\nabla^d \phi\Big),\\
        R^a_{\hphantom{a}bcd} &= e^{-\phi} R^a_{3\: bcd} - \frac{1}{2} e^{-\phi}\left(\delta_c^{\hphantom{c}a}\nabla_d \nabla_b \phi - \delta_d^{\hphantom{d}a}\nabla_c \nabla_b \phi - \eta_{cb}\nabla_d \nabla^a \phi + \eta_{db} \nabla_c \nabla^a \phi\right) \\
        &\hspace{0.5cm} + \frac{1}{4} e^{-\phi} \Big(\delta_c^{\hphantom{c}a}\nabla_d \phi \nabla_b \phi - \delta_d^{\hphantom{d}a}\nabla_C \phi \nabla_b \phi - \delta_c^{\hphantom{c}a}\eta_{bd}\nabla_e \phi \nabla^e \phi \\
        &\hspace{1.75cm} + \delta_d^{\hphantom{d}a}\eta_{bc}\nabla_e \phi \nabla^e \phi + \eta_{bd} \nabla^a \phi \nabla_c \phi - \eta_{bc}\nabla^a \phi \nabla_d \phi\Big)\\
        &\hspace{0.5cm} + \frac{1}{2} e^{-3\phi} \left(F^a_{\hphantom{a}b} F_{cd} + \frac{1}{2} F^a_{\hphantom{a}c} F_{bd} - \frac{1}{2} F^a_{\hphantom{a}d} F_{bc}\right).
    \end{aligned}
\end{equation}
From here, we can compute the Ricci tensor components:
\begin{equation}
    \begin{aligned}
        R_{00} &= -\frac{1}{2}e^{-\phi} \nabla_a \nabla^a \phi + \frac{1}{4} e^{-3\phi} F^{ab} F_{ab},\\
        R_{0a} &= \frac{1}{2}\nabla_b\left(e^{-2\phi} F^b_{\hphantom{b}a}\right),\\
        R_{ab} &= e^{-\phi}R_{3\: ab} - \frac{1}{2} e^{-\phi} \eta_{ab}\nabla_c \nabla^c \phi - \frac{1}{2}e^{-\phi}\nabla_a\phi \nabla_b \phi - \frac{1}{2}e^{-3\phi} F_{ac} F^c_{\hphantom{c}b}.
    \end{aligned}
\end{equation}
Finally, the Ricci scalar is
\begin{equation}
\label{eq:RicciScalarKK}
    R = e^{-\phi} R_3 - e^{-\phi} \nabla_a \nabla^a \phi - \frac{1}{2} e^{-\phi}\nabla_a \phi \nabla^a \phi + \frac{1}{4}e^{-3\phi} F^{ab} F_{ab}.
\end{equation}

We now proceed to the Weyl tensor, given by
\begin{equation}
    C_{\alpha\beta\gamma\delta} = R_{\alpha\beta\gamma\delta} + \frac{1}{2}\left(R_{\alpha\delta}\eta_{\beta\gamma} - R_{\alpha\gamma}\eta_{\beta\delta} + R_{\beta\gamma}\eta_{\alpha\delta} - R_{\beta\delta}\eta_{\alpha\gamma}\right) + \frac{R}{6}\left(\eta_{\alpha\gamma}\eta_{\beta\delta} - \eta_{\alpha\delta}\eta_{\beta\gamma}\right).
\end{equation}
The two components of the Weyl tensor of interest for our purposes are
\begin{equation}
    \begin{aligned}
        C_{0a0b} &= \frac{1}{2}e^{-\phi}\left(R_{3\: ab} - \frac{1}{3}R_3 \eta_{ab}\right) - \frac{1}{2}e^{-\phi}\left(\nabla_b \nabla_a \phi - \frac{1}{3}\eta_{ab}\nabla_c \nabla^c \phi\right) \\
        & \hspace{0.5cm} + \frac{1}{2} e^{-\phi}\left(\nabla_a \phi \nabla_b \phi - \frac{1}{3}\eta_{ab} \nabla_c \phi \nabla^c \phi\right) + \frac{1}{2}e^{-3\phi} \left(F_a^{\hphantom{a}c}F_{bc} - \frac{1}{3}\eta_{ab} F_{cd} F^{cd}\right),\\
        C_{0abc} &= -\frac{1}{2}e^{-2\phi}\left(\nabla_a F_{bc} + \frac{1}{2}\eta_{ac}\nabla_d F^d_{\hphantom{d}b} - \frac{1}{2}\eta_{ab}\nabla_d F^d_{\hphantom{d}c}\right) \\
        &\hspace{0.5cm} + \frac{1}{2}e^{-2\phi}\left(2F_{bc}\nabla_a \phi + F_{ac}\nabla_b \phi - F_{ab}\nabla_c \phi + \frac{3}{2}\left(F_{cd}\eta_{ab} - F_{bd}\eta_{ac}\right)\nabla^d \phi\right).
    \end{aligned}
\end{equation}

One last useful object to compute is the determinant, $\sqrt{-g} = \det\left(e^\alpha_{\hphantom{\alpha}\mu}\right)$, which can be done as follows
\begin{equation}
    \det\begin{pmatrix}
        e^{-\phi/2} & e^{\phi/2} A_i \\
        0 & \tilde{e}
    \end{pmatrix} = e^{\kappa_1 \phi} \det\left(\tilde{e}\right) = e^{-\phi/2} \sqrt{\tilde{g}} = e^{\phi}\sqrt{g_3}.
\end{equation}
Combining this with the form of the Ricci scalar in the Kaluza--Klein decomposition in \autoref{eq:RicciScalarKK}, the action, upon integrating out the total derivative and switching back to the coordinate basis, becomes 
\begin{equation}
\label{eq:KKAction}
    S = \int \dd[3]{x} \sqrt{g_3} \left(R_3 - \frac{1}{2}\nabla_i \phi \nabla^i \phi + \frac{1}{4} e^{-2\phi} F^{ij} F_{ij}\right).
\end{equation}

The last objects we will need are the electric and magnetic components of the Weyl tensor\footnote{A helpful introduction is on this \href{https://duetosymmetry.com/notes/notes-on-the-E-B-and-3+1-decomp-of-Riem/}{page} by Leo Stein.},
\begin{equation}
    E_{\alpha\beta} = C_{\alpha \gamma \beta \delta} n^\gamma n^\delta \qquad \text{and} \qquad B_{\alpha\beta} = \frac{1}{2}\epsilon_{\alpha \gamma}^{\hphantom{\alpha\gamma}\epsilon\zeta} C_{\epsilon\zeta\beta\delta} n^\gamma n^\delta,
\end{equation}
where $n^\alpha$ is the normal vector to the 3D hypersurface. In the tetrad basis, this is simply $n^0 = 1$ and $n^a = 0$. The the electric and magnetic parts of the Weyl tensor in the tetrad basis thus reduce to
\begin{equation}
    E_{ab} = C_{0a0b} \qquad \text{and} \qquad B_{ab} = \frac{1}{2}\epsilon_a^{\hphantom{a}cd} C_{0bcd}.
\end{equation}
Due to the form of the tetrad, this can be readily written in the coordinate basis. The electric tensor is
\begin{equation}
\label{eq:EWeyl}
\begin{aligned}
    E_{ij} &= \frac{1}{2}e^{-\phi}\left(R_{3\: ij} - \frac{1}{3} R_3 g_{3\: ij}\right) - \frac{1}{2}e^{-\phi}\left(\nabla_j \nabla_i \phi - \frac{1}{3}g_{ij} \nabla_k \nabla^k \phi\right) \\
    &\hspace{0.5cm} + \frac{1}{2}e^{-\phi}\left(\nabla_i \phi\nabla_j\phi - \frac{1}{3}g_{ij}\nabla_k \phi \nabla^k \phi\right) + \frac{1}{2}e^{-3\phi} \left(F_i^{\hphantom{i}k}F_{jk} - \frac{1}{3}g_{ij} F_{kl} F^{kl}\right).
\end{aligned}
\end{equation}
Meanwhile, the magnetic tensor in the coordinate basis is
\begin{equation}
\label{eq:BWeyl}
\begin{aligned}
    B_{ij} &= -\frac{1}{4}\epsilon_i^{\hphantom{i}kl} e^{-2\phi}\Big(\nabla_j F_{kl} + \frac{1}{2}g_{3\:jl}\nabla_m F^m_{\hphantom{m}k} - \frac{1}{2} g_{3\:jk} \nabla_m F^m_{\hphantom{m}k} - 2F_{kl} \nabla_j \phi \\
    &\hspace{2cm} - F_{jl}\nabla_k \phi + F_{jk}\nabla_l\phi - \frac{3}{2}\left(F_{lm}g_{3\:jk} - F_{km}g_{3\:jl}\right)\nabla^m\phi\Big).
\end{aligned}
\end{equation}
Note that both of these tensors are symmetric and traceless. As such, they each contain 5 independent degrees of freedom, which in total capture the 10 degrees of freedom the Weyl tensor had.

\section{Gauge-Independent Variables}
\label{a:GIVariables}
In this section, we study how the Kaluza--Klein variables, defined in \autoref{eq:KKMetric}, transform under gauge transformations. Throughout this section the bar notation denotes a quantity with respect to the background metric. To begin, consider metric perturbations on a Schwarzschild background,
\begin{equation}
g_{\mu\nu}=\bar{g}_{\mu\nu}+\delta g_{\mu\nu},
\end{equation}
where $\bar{g}_{\mu\nu}$ is the background Schwarzschild metric. We will also take the Kaluza--Klein parameters and vary them in the same way,
\begin{equation}
    \phi = \bar{\phi} + \delta \phi, \qquad A_i = \bar{A}_i + \delta A_i, \qquad \text{and} \qquad g_{3\: ij} = \bar{g}_{3\:ij} + h_{ij}.
\end{equation}
The background values of the Kaluza--Klein metric are given in \autoref{eq:BackgroundEqns} and reproduced below for convenience
\begin{equation}
    e^{\bar{\phi}} = \frac{r^2}{\Delta}, \qquad \bar{A}_i = 0, \qquad \text{and} \qquad \dd{\bar{s}}^2_3 = \dd{r}^2 + \Delta \dd{\Omega}^2,
\end{equation}
where $\dd{\bar{s}}^2_3$ is the line element associated to $\bar{g}_{3\: ij}$. Similarly, $\delta g_{\mu\nu}$ is related to the perturbed Kaluza--Klein variables by
\begin{equation}
\label{eq:PerturbMetricToKK}
    \delta g_{tt} = \frac{\Delta}{r^{2}}\delta\phi, \qquad
    \delta g_{ti} = -\frac{\Delta}{r^{2}}\delta A_{i}, \qquad \text{and} \qquad 
    \delta g_{ij} = \frac{r^{2}}{\Delta}\bar{g}_{3\: ij}\delta\phi+\frac{r^{2}}{\Delta}h_{ij}.
\end{equation}

Under gauge transformation, $\delta g_{\mu\nu}$ transforms as
\begin{equation}
\delta g_{\mu\nu} \rightarrow \delta g_{\mu\nu}' = \delta g_{\mu\nu} + \overline{\nabla}_{\mu}\xi_{\nu} + \overline{\nabla}_{\nu}\xi_{\mu},
\end{equation}
where $\xi_{\mu}$ is a diffeomorphism. The individual metric components transform as:
\begin{equation}
    \begin{aligned}
        \delta g_{tt}' &= \delta g_{tt} - \frac{r_s \Delta}{r^4}\xi_r, \qquad & \qquad \delta g_{rr}' &= \delta g_{rr} + 2\partial_r \xi_r + \frac{r_s}{\Delta}\xi_r,\\
        \delta g_{tr}' &= \delta g_{tr} + \partial_r \xi_t - \frac{r_s}{\Delta}\xi_t, & \delta g_{ra}' &= \delta g_{ra} + \partial_r \xi_{a} + D_{a} \xi_r - \frac{2}{r}\xi_{a},\\
        \delta g_{ta}' &= \delta g_{ta} + D_{a} \xi_t, & \delta g_{ab}' &= \delta g_{ab} + D_{a} \xi_{b} + D_{b}\xi_{a} + \frac{2\Delta}{r} \gamma_{ab}\xi_r,
    \end{aligned}
\end{equation}
where $D_{a}$ is the covariant derivative on the 2-sphere compatible with $\gamma_{ab}$. We can now recast these in terms of the perturbed Kaluza--Klein variables using \autoref{eq:PerturbMetricToKK}:
\begin{equation}
\label{eq:KKGaugeTrans}
    \begin{aligned}
        \delta \phi' &= \delta \phi - \frac{r_s}{r^2}\xi_r, \qquad & \qquad  h_{ra}' &= h_{ra} + \frac{\Delta}{r^2}\partial_r \xi_{a} + \frac{\Delta}{r^2} D_{a}\xi_r - \frac{2\Delta}{r^3}\xi_{a}\\
        \delta A_i' &= \delta A_i - \partial_i\left(\frac{r^2}{\Delta}\xi_t\right), & h_{ab}' &= h_{ab} + \frac{\Delta}{r^2}D_{a} \xi_{b} + \frac{\Delta}{r^2} D_{b}\xi_{a} + \frac{\Delta \Delta'}{r^2}\gamma_{ab}\xi_r\\
        h_{rr}' &= h_{rr} + \frac{2\Delta}{r^2}\partial_r \xi_r + \frac{2 r_s}{r^2}\xi_r,\\
    \end{aligned}
\end{equation}
Note that $A_i$ uses a full spatial index $i$. This gauge transformation reduces to the usual gauge transformation on the vector potential. Since $\xi_t$ only appears in the transformation of $A_i$, this can be treated separately. Indeed, in \Cref{s:Spin2}, we see that the part of the action that contains $A_i$ completely decouples, leaving behind an action involving $\delta \phi$ and $h_{ij}$. We will restrict our attention to those variables for the remainder of the section. 

There still remains a vector, $h_{ra}$, and tensor $h_{ab}$ that can be further decomposed. We decompose the vector piece to
\begin{equation}
        h_{ra} = D_{a}\mathcal{H}_E + \epsilon_{a}^{\hphantom{a}b} D_{a} \mathcal{H}_O,
\end{equation}
where, by construction, $\mathcal{H}_E$ is parity-even and $\mathcal{H}_O$ is parity-odd. For the tensor piece we first decompose into a trace and a trace-free part and then further decompose the trace-free part into an even and odd parity piece. Explicitly, 
\begin{equation}
        h_{ab} = \frac{1}{2}\gamma_{ab} K + \left(D_{a} D_{b} - \frac{1}{2}\gamma_{ab} D^2\right)\mathcal{G}_E + \frac{1}{2}\left(\epsilon_{a}^{\hphantom{a}c} D_{c} D_{b} + \epsilon_{b}^{\hphantom{b}c} D_{c} D_{a}\right)\mathcal{G}_O,
\end{equation}
where $K$ and $\mathcal{G}_E$ are the even-parity pieces and $\mathcal{G}_O$ is odd-parity. The even- and odd-parity sectors completely decouple in the equations of motion, so we additionally decompose the diffeomorphism as
\begin{equation}
\xi_{\mu} = \left(\xi_t, \xi_r, \xi_{a}\right),
\end{equation}
where $\xi_t$ is entirely contained in the transformation of $\delta A_i$ and
\begin{equation}
\xi_{a} \equiv D_{a} \xi_E + \epsilon_{a}^{\hphantom{a}b} D_{b} \xi_O.
\end{equation}
Inserting this decomposition into the gauge transformations shown in \autoref{eq:KKGaugeTrans}, we find that the gauge transformations of these Kaluza--Klein variables become
\begin{equation}
\label{eq:FinalGaugeTrans}
    \begin{aligned}
        \delta\phi' &= \delta \phi - \frac{r_s}{r^2}\xi_r, \qquad & \qquad K' &= K + \frac{2\Delta}{r^2} D^2 \xi_E + \frac{2\Delta\Delta'}{r^2}\xi_r,\\
        h_{rr}' &= h_{rr} + \frac{2\Delta}{r^2} \partial_r \xi_r + \frac{2r_s}{r^2}\xi_r, & \mathcal{G}_E' &= \mathcal{G}_E + \frac{2\Delta}{r^2}\xi_E,\\
        \mathcal{H}_E' &= \mathcal{H}_E + \Delta\partial_r\left(\frac{1}{r^2}\xi_E\right) + \frac{\Delta}{r^2} \xi_r, & \mathcal{G}_O' &= \mathcal{G}_O + \frac{2\Delta}{r^2} \xi_O.\\
        \mathcal{H}_O' &= \mathcal{H}_O + \Delta\partial_r\left(\frac{1}{r^2}\xi_O\right),
    \end{aligned}
\end{equation}

Inspired by \cite{Martel2005}, we find it useful to build gauge-invariant combinations of the decomposed Kaluza--Klein variables, i.e. take combinations in such a way that all gauge dependence cancels. In the odd-parity section, there is one gauge independent variable
\begin{equation}
    \label{eq:OddGIV}
    \tilde{\mathcal{H}}_O = \mathcal{H}_O - \frac{\Delta}{2}\partial_r\left(\frac{1}{\Delta}\mathcal{G}_O\right).
\end{equation}
For the even parity sector, there is no unique choice of combinations. A convenient set of three gauge independent variables are:
\begin{equation}
    \label{eq:EvenGIV}
    \begin{aligned}
        \delta\tilde{\phi} &= \delta\phi - \frac{r_s}{\Delta}\mathcal{H}_E + \frac{r_s}{2}\partial_r\left(\frac{1}{\Delta}\mathcal{G}_E\right),\\
        \tilde{K} &= K - 2\Delta'\mathcal{H}_E + \Delta \Delta' \partial_r\left(\frac{1}{\Delta}\mathcal{G}_E\right) - D^2\mathcal{G}_E,\\
        \tilde{h}_{rr} &= h_{rr} - 2\partial_r\mathcal{H}_E + \partial_r\left(\Delta\partial_r\left(\frac{1}{\Delta}\mathcal{G}_E\right)\right).
    \end{aligned}
\end{equation}

\section{Chandrasekhar Duality of spin 2 Perturbations}
\label{a:Chandrasekhar}
Recall that the total spin 2 action with $\Omega = r^4/\Delta^2$ is \autoref{eq:CombinedSpin2Action}
\begin{equation}
S = -\frac{1}{2} \int \dd[3]{x} \sqrt{\hat{g}} \left(\hat{g}^{ij} \partial_i \phi \partial_j \phi + 2 \hat{R} \phi^2 + \Omega \hat{g}^{ij}  \partial_i \chi \partial_j \chi\right).
\end{equation}
For ease of notation we denote the even perturbation in this appendix by $\phi$ instead of $\delta \phi$. In addition to invariance under the melodic CKV symmetries that form an SO(3,1) group, there is an additional SO(2) symmetry that rotates the even to the odd sectors. To see this, we begin by discussing a simpler \textit{discrete} symmetry that maps even to odd perturbations and vice versa:
\begin{equation}
\phi \rightarrow - \Omega^{1/2} \chi \qquad \text{and} \qquad \chi \rightarrow \Omega^{-1/2} \phi. 
\end{equation}
The even action transforms as
\begin{subequations}
\begin{align}
S_{\text{even}} &\equiv -\frac{1}{2} \int \dd[3]{x} \sqrt{\hat{g}} \left(\hat{g}^{ij}\partial_i \phi \partial_j \phi + 2\hat{R} \phi^2\right),\\
&\rightarrow -\frac{1}{2} \int \dd[3]{x} \sqrt{\hat{g}} \left(\Omega \hat{g}^{ij} \partial_i \chi \partial_j \chi + \chi \partial_i \chi \partial^i \Omega + \frac{\chi^2}{4\Omega} \partial_i \Omega \partial^i \Omega + 2\Omega \hat{R} \chi^2\right),\\
&= -\frac{1}{2} \int \dd[3]{x} \sqrt{\hat{g}}\left(\Omega \hat{g}^{ij} \partial_i \chi \partial_j \chi + \left(2\hat{R} \Omega - \frac{1}{2} \Box \Omega + \frac{1}{4\Omega} \partial_i\Omega \partial^i \Omega\right) \chi^2 \right),\\
&= -\frac{1}{2}\int \dd[3]{x} \sqrt{\hat{g}} \: \Omega \hat{g}^{ij} \partial_i \chi \partial_j \chi.
\end{align}
\end{subequations}
We implemented the transformation in going from the first to the second line, utilizing the fact that $\partial_i \Omega^{1/2} = (1/2)\Omega^{-1/2} \partial_i \Omega$. In going from the second to the third line, we noted that $2\chi \partial_i \chi = \partial_i \chi^2$ and integrated by parts. Then, in going to the last line, we noted that the Ricci scalar is related to the scale factor $\Omega$ by 
\begin{equation}
2\hat{R} = \frac{1}{2 \Omega} \Box \Omega - \frac{1}{4\Omega^2} \partial_i \Omega \partial^i \Omega,
\end{equation}
where, as a reminder, the Ricci scalar $\hat{R}$ is equivalent to that in \autoref{eq:StaticScalarRicci}. Note that we have dropped all total derivative terms. Similarly, the odd metric transforms as
\begin{subequations}
\begin{align}
S_{\text{odd}} &\equiv - \frac{1}{2} \int \dd[3]{x} \sqrt{\hat{g}} \: \Omega \hat{g}^{ij} \partial_i \chi \partial_j \chi,\\
&\rightarrow -\frac{1}{2} \int \dd[3]{x} \sqrt{\hat{g}} \left(\hat{g}^{ij} \partial_i \chi \partial_j \phi - \frac{\phi}{\Omega} \partial_i \phi \partial^i \Omega + \frac{\phi^2}{4\Omega^2} \partial_i \Omega \partial^i \Omega\right),\\
&= -\frac{1}{2} \int \dd[3]{x} \sqrt{\hat{g}} \left(\hat{g}^{ij} \partial_i \phi \partial_j \phi + 2 \hat{R}\phi^2\right). 
\end{align}
\end{subequations}
In going from the first to the second line, we implemented the transformation noting that $\partial_i \Omega^{-1/2} = -(1/2)\Omega^{-3/2} \partial_i \Omega$. In going to the last line, we noted that $2\phi \partial_i \phi = \partial_i \phi^2$ and integrated by parts. The remaining term, as pointed out in the transformation of the even action, is equal to a Ricci scalar term. Therefore, the even and odd spin 2 perturbations transform into one another under a simple re-scaling. 

The observation that there is a field-space symmetry that transforms even-parity perturbations into odd-parity perturbations is not new. As reviewed and explored in \cite{Solomon:2023} there exists a hidden off-shell symmetry of the action, called the \textit{Chandrasekhar duality}, that maps between the odd--parity Regge--Wheeler variable and the even--parity Zerilli variable. In the case at hand, as claimed before, this local symmetry is an SO(2) rotation in field space: 
\begin{equation}
\begin{pmatrix}
\phi \\ \chi 
\end{pmatrix} \rightarrow \begin{pmatrix} \cos\alpha & -\Omega^{1/2} \sin\alpha \\
\Omega^{-1/2} \sin\alpha & \cos\alpha 
\end{pmatrix} \begin{pmatrix} \phi \\ \chi \end{pmatrix}.
\end{equation}
We will show that this is a local symmetry of the action by working with the infinitesimal transformations,
\begin{equation}
\phi \rightarrow \phi - \Omega^{1/2} \alpha \chi \qquad \text{and} \qquad \chi \rightarrow \chi + \Omega^{-1/2} \alpha \phi.
\end{equation}
The variation of the action is 
\begin{equation}
\delta S = \int \dd[3]{x} \sqrt{\hat{g}} \: \alpha \left[\partial_i \phi \partial^i \left(\Omega^{1/2} \chi\right) + 2 \hat{R} \phi \chi \Omega^{1/2} - \Omega \partial_i \chi \partial^i\left(\Omega^{-1/2} \phi\right)\right].
\end{equation}
By explicit calculation, we find
\begin{equation}
\partial_i \phi \partial^i \left(\Omega^{1/2} \chi\right) + 2 \hat{R} \phi \chi \Omega^{1/2} - \Omega \partial_i \chi \partial^i\left(\Omega^{-1/2} \phi\right) = \nabla_i\left(\phi \chi \nabla^i \Omega^{1/2}\right),
\end{equation}
where the covariant derivative with associated to the metric $\hat{g}_{ij}$. Therefore, the action indeed changes by a total derivative given by the right-hand side of the above equation. By the general formula for the Noether current \autoref{eq:GeneralNoetherCurrent}, we find that the Noether current associated to this field space SO(2) symmetry is 
\begin{equation}
J^i = \Omega^{1/2}\left(\chi \nabla^i \phi - \nabla^i \chi\right) - \phi \chi \nabla^i \Omega^{1/2}. 
\end{equation}
In the limit where $\Omega\rightarrow 1$, this reduces to the familiar SO(2) rotation Noether current, which is also the current for electric-magnetic duality. 

\section{Connecting to Higher Symmetries via the Spin Ladder}
\label{s:HigherSpinLadder}

In \Cref{s:SchwarzschildVector}, we found that the fields $A_t$ and $\chi$ lived in an effective three-dimensional space, allowing us to construct symmetries of their equations of motion out of the melodic CKVs. These symmetries were distinct from the symmetries of the spin 0 equation of motion, as the effective three-dimensional space was different, though, as discussed, the particular conformal relation between the two spaces let us use the same melodic CKVs in both cases. 

A more direct approach to finding symmetries of the equations of motion for higher spin fields would be to use the spin ladder operators defined in \cite{Hui2021}, which take solutions of the static Teukolsky equation with spin $s$ to solutions with spin $s\pm 1$. The Teukolsky equation is satisfied by the fields $\Phi^{(s)}(r,\theta,\phi)$, which for $s\neq0$ are Newman--Penrose scalars constructed using contractions of a null tetrad with a field with $|s|$ indices. 

After a decomposition over spin-weighted spherical harmonics, the field becomes
\begin{align}
    \Phi^{(s)}(r,\theta,\phi) = \sum_{\ell = |s|}^{\infty} \sum_{m
  = -\ell}^{\ell} c_{\ell m} \phi_{\ell m}^{(s)}(r) Y_{\ell m}^{(s)}(\theta,\phi).
\end{align}
The spin ladder operators defined in \cite{Hui2021} are 
\begin{equation}
\label{eq:SpinLadder}
    E_s^+ = \Delta \partial_r - s \Delta' = \Delta^{s+1} \partial_r \Delta^{-s},
    \qquad
    E_s^- = \partial_r.
\end{equation}
These act on the radial fields $\psi_\ell^{(s)} = \Delta^s \phi_\ell^{(s)}$, where we have dropped the $m$ subscript. We also have the angular spin operators
\begin{subequations}
\begin{align} 
    \eth_s &= -\left(\sin\theta\right)^s \left(\partial_\theta + i \csc\theta\partial_\varphi\right)\left(\sin\theta\right)^{-s}\\
    &= -\left(\partial_\theta + i\csc\theta\partial_\phi - s\cot\theta\right)\\
    \overline{\eth}_s &= -\left(\sin\theta\right)^{-s} \left(\partial_\theta - i\csc\theta\partial_\varphi\right)\left(\sin\theta\right)^s\\
    &= -\left(\partial_\theta - i\csc\theta\partial_\phi + s\cot\theta\right),
\end{align}
\end{subequations}
which have the effect of raising and lowering the spin of the spin-weighted spherical harmonics:
\begin{align}
    \eth_s Y^{(s)}_{\ell m} = 
    \sqrt{(\ell-s)(\ell+s+1)} Y^{(s+1)}_{\ell m}, \quad
    \overline{\eth}_s Y^{(s)}_{\ell m} = 
    -\sqrt{(\ell+s)(\ell-s+1)} Y^{(s-1)}_{\ell m}.
\end{align}
These harmonics are conveniently defined via these operators in terms of the usual spherical harmonics:
\begin{align}
    Y^{(s)}_{\ell m} =
    \begin{cases}
    \sqrt{\frac{(\ell-s)!}{(\ell+s)!}} 
    \eth_{s-1} \cdots \eth_0 Y_{\ell m}, 
    & 0 \leq s \leq \ell\\
    (-1)^s \sqrt{\frac{(\ell+s)!}{(\ell-s)!}} 
    \overline{\eth}_{|s|-1} \cdots \eth_0 Y_{\ell m}, 
    & -\ell \leq s \leq 0\\
    0, & \ell < |s|
    \end{cases}.
\end{align}
These operators also obey the identities
\begin{subequations}
\begin{align} 
    \overline{\eth}_{s+1} \eth_s  
    &= \nabla^2_{S^2} + 2is\cot\theta\csc\theta\,\partial_\varphi -s\left(s\cot^2\theta-1\right),\\
    \eth_{s-1} \overline{\eth}_s &= \nabla^2_{S^2} + 2is\cot\theta\csc\theta\,\partial_\varphi -s\left(s\cot^2\theta+1\right).
\end{align}
\end{subequations}

With all this machinery in place, we can now define the ``spin ladder symmetry''
\begin{align}
\label{eq:SpinLadderSymmetry}
    \delta^{\text{ladder}}_\xi \Phi^{(s)} = \Delta^{-s} 
    E_{s-1}^+ \eth_{s-1} E_{s-2}^+ \eth_{s-2} 
    \cdots E_1^+ \eth_1 E_0^+ \eth_0
    \delta_\xi E_1^- \overline{\eth}_1 E_2^- \overline{\eth}_2 
    \cdots E_{s-1}^- \overline{\eth}_{s-1} E_s^- \overline{\eth}_s \Delta^s \Phi^{(s)},
\end{align}
where 
\begin{align}
    \delta_\xi \Phi^{(0)} = \left(\xi^i \partial_i + \frac{1}{6}\left(\nabla \cdot \xi\right)\right)\Phi^{(0)},
\end{align}
and $\xi$ is a melodic CKV. Here every operator acts on everything to its right. The idea is to use repeated applications of $E_s^-$ and $\overline{\eth}_s$ to lower the spin of $\Phi^{(s)}$ down to $s = 0$, apply the scalar symmetry operator $\delta_\xi$, and then raise the spin back up to $s$ by repeated applications of $E_s^+$ and $\eth_s$. By construction, this process will map a solution $\Phi^{(s)}$ of the static Teukolsky equation to another solution. 

Of course, the operator defined in \autoref{eq:SpinLadderSymmetry} is quite cumbersome. It has $4s+1$ derivatives, so even in the $s=1$ case it cannot be simply related to the spin 1 symmetries found in \Cref{s:SchwarzschildVector}. However, as we shall show, the $s=1$ spin ladder symmetry is in fact connected to the previous spin 1 symmetries.

For $s = 1$, \autoref{eq:SpinLadderSymmetry} becomes
\begin{align}
    \delta^{\text{ladder}}_\xi \Phi^{(1)} = \Delta^{-1} E_0^+ \eth_0 
    \delta_\xi E_1^- \overline{\eth}_1 \Delta \Phi^{(1)}.
\end{align}
Focusing on the electrostatic case, we can express the Newman-Penrose scalar $\Phi^{(1)} = F_{\mu\nu}l^\mu m^\nu$ in terms of $A_0$:
\begin{align}
    \Phi^{(1)} = \frac{1}{\sqrt{2}}\frac{r}{\Delta}\eth_0 A_0.
\end{align}
We can then write
\begin{subequations}
\begin{align}
    \delta^{\text{ladder}}_\xi \Phi^{(1)} 
    &= \frac{1}{\sqrt{2}} \Delta^{-1} E_0^+ \eth_0 \delta_\xi E_1^- \overline{\eth}_1 r \eth_0 A_0\\
    &= \frac{1}{\sqrt{2}} \partial_r \eth_0 \delta_\xi \partial_r r\overline{\eth}_1 \eth_0 A_0\\
    &= \frac{1}{\sqrt{2}} \partial_r \eth_0 \delta_\xi \partial_r r \nabla^2_{S^2} A_0.
\end{align}
\end{subequations}
We let $\tilde{\delta}_\xi$ denote the symmetry for the scalar $A_0$ derived in \Cref{s:SchwarzschildVector}, i.e.,
\begin{align}
    \tilde{\delta}_\xi A_0 = \left(\xi^i \partial_i + \frac{1}{6}\tilde{\nabla} \cdot \xi\right) A_0,
\end{align}
where the tilde on the covariant derivative indicates that it is the covariant derivative associated with the the effective metric in \autoref{eq:3DSchwarzschildVectorEffectiveMetric}. We now ask if $\delta^{\text{ladder}}_\xi \Phi^{(1)}$ is related to this symmetry. We focus on the electrostatic case, where the Newman-Penrose scalar $\Phi^{(1)} = F_{\mu\nu}l^\mu m^\nu$ can be expressed as
\begin{align}
    \Phi^{(1)} = \frac{1}{\sqrt{2}} \frac{r}{\Delta}\eth_0 A_0.
\end{align}
By using the conformal Killing equation, the melodic condition \autoref{eq:MelodicCondition}, and \cite[eq. (F.3)]{Berens2023}, we can show that a melodic CKV in our space must obey
\begin{align}
\label{eq:DivergenceIdentity}
    \nabla \cdot \xi = 3\partial_r \xi^r = \frac{3\Delta'}{\Delta} \xi^r,
\end{align}
which shall be useful in the following simplifications. After a heroic amount of algebra and frequent use of \autoref{eq:DivergenceIdentity}, we find
\begin{align}
\label{eq:DeltaLadderFinal}
    \delta^{\text{ladder}}_\xi \Phi^{(1)} 
    &= \frac{1}{\sqrt{2}} \frac{r}{\Delta}
    \eth_0\bigg[-\tilde{\delta}_\xi \nabla^2_{S^2} \nabla^2_{S^2} A_0
    + \delta_\text{new}\nabla_{S^2}^2 A_0
    \notag
    \\&\quad
    + \frac{r^4}{\Delta} \xi^i \partial_i \nabla_{S^2}^2 \tilde{\Box} A_0
    + \frac{r^4}{2\Delta}\left(\frac{10}{r}-\frac{\Delta'}{\Delta}\right)\xi^r \nabla_{S^2}^2 \tilde{\Box}A_0\bigg],
\end{align}
where 
\begin{align}
\label{eq:NewSymmetry}
    \delta_\text{new} &= \Delta\left[\left(\partial_r \xi^\theta\right)\partial_\theta + \left(\partial_r \xi^\varphi\right)\partial_\varphi\right]\partial_r
    + \frac{\Delta}{r}\left[\left(\partial_r \xi^\theta\right)\partial_\theta + \left(\partial_r \xi^\varphi\right)\partial_\varphi\right]
    \notag
    \\&\quad
    + \xi^r \partial_r - \frac{\Delta'}{2\Delta} \xi^r \nabla_{S^2}^2 + \frac{1}{r} \xi^r.
\end{align}
We see that the two final terms inside the the square brackets in \autoref{eq:DeltaLadderFinal} vanish when $A_0$ satisfies its equation of motion. 
Since $\delta^{\text{ladder}}_\xi \Phi^{(1)}$ solves the $s=1$ Teukolsky equation by construction, the remaining terms in the square brackets must solve the equation of motion for $A_0$. The first term is just the symmetry from \Cref{s:SchwarzschildVector} composed with $\left(\nabla_{S^2}^2\right)^2$. Since $\nabla_{S^2}^2$ is also a symmetry of the equation of motion, this term will solve the equation of motion for $A_0$ on its own. It follows that the second term must do so as well, and thus $\delta_\text{new}$ must also be a symmetry of the equation of motion for $A_0$. Note that, though not apparent from the form of \autoref{eq:NewSymmetry}, if $\xi$ is an ordinary Killing vector, one can use the Killing equation to show that
\begin{gather}
    \xi^r = 0, \qquad \partial_r \xi^\theta = 0, \qquad \partial_r \xi^\varphi = 0,
\end{gather}
so $\delta_\text{new}$ vanishes. 

This new second-order symmetry of the equation of motion for $A_0$ is not obviously expressible in terms of $\tilde{\delta}_\xi$ or other previous symmetries, but we know that any second-order symmetry of the Laplacian can be associated with a conformal Killing tensor. The three-dimensional metric in \autoref{eq:3DSchwarzschildVectorEffectiveMetric} is conformally flat, which further implies that any conformal Killing tensor is reducible, i.e., expressible as a linear combination of symmetric outer products of CKVs. More detail on conformal Killing tensors and symmetries of the Laplacian can be found below in \Cref{subsec:CKTSymmetries}. For now, we simply state that if the melodic CKV $\xi$ takes the form
\begin{align}
    \xi = \sum_{a=1}^3 c_a K_a
\end{align}
for some coefficients $c_a$ and with the $K_a$ given in \autoref{eq:AllMCKVs1} -- \autoref{eq:AllMCKVs3}, then the new symmetry $\delta_\text{new}$ can be associated with the following conformal Killing tensor:
\begin{align} 
\label{eq:NewSymmetryCKT}
    T^{ij} &= -c_a \epsilon^{abc} J_b^{(i} K_c^{j)},
\end{align}
where the $J_a$ are also given in \autoref{eq:AllMCKVs4} -- \autoref{eq:AllMCKVs6}. More precisely, $\delta_\text{new} = \delta_T$, where the form of $\delta_T$ is given in the next section.

\subsection{Conformal Killing Tensors and Symmetries of the Laplacian}
\label{subsec:CKTSymmetries}

On a manifold of dimension $d$, a two-index (not necessarily traceless) conformal Killing tensor obeys
\begin{align} 
\label{eq:CKTEquation}
    \nabla^{(\mu} T^{\nu\rho)} = \frac{2}{d+2} \left(g^{(\mu\nu} \nabla_\sigma T^{\rho)\sigma}
    +\frac{1}{2} g^{(\mu\nu} \nabla^{\rho)} {T^\sigma}_\sigma
    \right).
\end{align}
On a conformally flat manifold, a conformal Killing tensor corresponds to a unique second-order symmetry of the conformal Laplacian
\begin{align}
\label{eq:ConformalLaplacian}
    \Box_c = \Box - \xi_c R, \qquad \xi_c = \frac{d-2}{4(d-1)}.
\end{align}
i.e., there exist unique second-order operators $\delta_T$ and $\overline{\delta}_{T}$ which obey\footnote{If the manifold is not conformally flat, the right hand side of \autoref{eq:CKTSymmetry} instead involves the obstruction vector defined in \cite{Michel2014}. This vector vanishes if the manifold is conformally flat, though the converse is not true.}
\begin{align}
\label{eq:CKTSymmetry}
    \left(\Box_c\delta_T - \overline{\delta}_T\Box_c\right) \Phi &=0.
\end{align}
These unique operators are\footnote{To our knowledge, this is a new result, as previous work either uses a traceless conformal Killing tensor or an ordinary Killing tensor. This result generalizes those cases, as every ordinary Killing tensor is trivially a conformal Killing tensor, but not necessarily a traceless one.}
\begin{subequations}
\label{eq:CKTSymmetryOperators}
\begin{align}
     \delta_T\Phi &= T^{\mu\nu}\nabla_\mu\nabla_\nu\Phi 
     + \frac{d}{d+2}\nabla_\mu T^{\mu\nu}\nabla_\nu \Phi 
     + \frac{d(d-2)}{4(d+1)(d+2)}\nabla_\mu\nabla_\nu T^{\mu\nu} \Phi 
     \\&\quad\notag
     - \frac{d-2}{4(d+1)(d+2)}\Box {T^\mu}_\mu \Phi 
     - \frac{1}{d+2} \nabla_\nu {T^\mu}_\mu\nabla^\nu\Phi
     - \frac{d+2}{4(d+1)} R_{\mu\nu} T^{\mu\nu} \Phi,
     \\&\quad\notag
     + \frac{1}{2(d-1)(d+1)} R {T^\mu}_\mu\Phi,\\
    \overline{\delta}_T\Phi &= T^{\mu\nu}\nabla_\mu\nabla_\nu \Phi 
    + \frac{d+4}{d+2}\nabla_\mu T^{\mu\nu}\nabla_\nu\Phi
    + \frac{d+4}{4(d+1)}\nabla_\mu\nabla_\nu T^{\mu\nu}\Phi 
    \\&\quad\notag
    + \frac{3}{4(d+1)}\Box {T^\mu}_\mu \Phi 
    + \frac{1}{d+2} \nabla_\nu {T^\mu}_\mu \nabla^\nu \Phi 
    - \frac{d+2}{4(d+1)} R_{\mu\nu} T^{\mu\nu} \Phi
    \\&\quad\notag
    + \frac{1}{2(d-1)(d+1)} R {T^\mu}_\mu \Phi.
\end{align}
\end{subequations}
Of course, we are interested in symmetries of the Laplacian, not the conformal Laplacian. If we replace the parameter $\xi_c$ in \autoref{eq:ConformalLaplacian} with a generic parameter $\xi$, which includes the case of $\xi=0$ for the usual Laplacian, we can show that on a conformally flat manifold,
\begin{align}
    \label{eq:RicciLaplacianCKT}
    \left[\left(\Box-\xi R\right) \delta_T - \overline{\delta}_T \left(\Box-\xi R\right)\right]\Phi
    &= \frac{2\left(\xi-\xi_c\right)}{d+2}\bigg\{\bigg[
    2R\nabla_\nu T^{\mu\nu}
    + R\nabla^\mu {T^\nu}_\nu
    + (d+2) \nabla_\nu R \, T^{\mu\nu} \bigg]\nabla_\mu \Phi
    \notag\\&\phantom{=}
    + \frac{1}{2}\nabla_\mu\bigg[2R\nabla_\nu T^{\mu\nu}
    + R\nabla^\mu {T^\nu}_\nu
    + (d+2) \nabla_\nu R \, T^{\mu\nu}\bigg] \Phi
    \bigg\}.
\end{align}
The vanishing of the right hand side of this expression is the analogue of the \autoref{eq:MelodicCondition} for conformal Killing tensors (when working on a conformally flat manifold).\footnote{As in the case of CKVs, we can show that all conformal Killing tensors are melodic in flat space. Likewise, in a non-flat maximally symmetric space, i.e., one with $\partial_\mu R = 0$ but $R \neq0$, the only melodic conformal Killing tensors are ordinary Killing tensors. Furthermore, when $T^{\mu\nu}$ is a reducible conformal Killing tensor, it can be shown that it is a melodic conformal Killing tensor if it is composed of symmetric outer products of melodic conformal Killing vectors.} When the quantity in square brackets in \autoref{eq:RicciLaplacianCKT} vanishes, $\delta_T$ is a symmetry of the generic Ricci-coupled Laplacian. We can check that the reducible conformal Killing tensor defined in \autoref{eq:NewSymmetryCKT} is melodic, i.e., it makes that quantity vanish, and thus $\delta_\text{new}$ is a symmetry of the Laplacian.

\newpage


\addcontentsline{toc}{section}{References}
\bibliographystyle{JHEP}
\bibliography{Bibliography.bib}

\end{document}